\documentclass[12pt]{article}
\usepackage{xr}
\makeatletter
\newcommand*{\addFileDependency}[1]{
	\typeout{(#1)}
	\@addtofilelist{#1}
	\IfFileExists{#1}{}{\typeout{No file #1.}}
}
\makeatother

\newcommand*{\myexternaldocument}[2]{%
	\externaldocument[#2]{#1}%
	\addFileDependency{#1.tex}%
	\addFileDependency{#1.aux}%
}
\myexternaldocument{proof-Sep-02-2017}{}
\usepackage{amsthm}
\usepackage{booktabs}
\usepackage{amsmath}
\usepackage{multirow}

\usepackage{color}
\usepackage{graphicx}
\usepackage{amsmath, amssymb, amsfonts, amsthm}
\usepackage{cite}
\usepackage{graphics}
\usepackage{color}
\usepackage{epstopdf}
\usepackage{caption}
\usepackage{subcaption}
\usepackage{pdfsync}
\usepackage{pdfpages}
\usepackage[authoryear]{natbib}
\usepackage{hyperref}
\usepackage[usenames,dvipsnames]{xcolor}

\newcommand{\field}[1]{\mathbb{#1}}
\newcommand{\R}{\field{R}}

\newcommand{\p}{\field{P}}

\newcommand{\E}{\field{E}}

\newcommand{\FF}{\mathcal{F}}

\def\argmax{\mathop{\mbox{argmax}}}
\def\argmin{\mathop{\mbox{argmin}}}
\theoremstyle{Conjecture}
\theoremstyle{example}
\theoremstyle{remark}
\theoremstyle{lemma}
\theoremstyle{definition}
\theoremstyle{proposition}
\theoremstyle{condition}
\newtheorem{theorem}{Theorem}[section]
\newtheorem{algorithm}{Algorithm}[section]

\newtheorem{example}{Example}[section]
\newtheorem{remark}{Remark}[section]
\newtheorem{lemma}{Lemma}[section]
\newtheorem{definition}{Definition}[section]

\def\cod{\stackrel{\cal D}{\longrightarrow}}

\def\lf{\lfloor}
\def\rf{\rfloor}

\usepackage{geometry}
\usepackage{setspace}
\begin{document}

	\centerline{\bf Change Point Analysis of Correlation in Non-stationary Time Series}
	\bigskip
	\font\n=cmcsc10
	\centerline{\sc Holger Dette, Weichi Wu\footnote{Corresponding author.
			Department of Statistics, University College, Gower Street
			London, WC1E 6BT, UK
			\newline
			\em E-mail: w.wu@ucl.ac.uk}
		and Zhou Zhou}
	\ \\
	\centerline{\sc \small Ruhr-Universit\"at Bochum, University College London and University of Toronto}
	
	\begin{abstract}{
			A restrictive assumption in change point analysis
			is ``stationarity under the null hypothesis of  no change-point'',
			which  is crucial for asymptotic theory   but
			not very realistic from a practical point of view. For example, if
			change point analysis for  correlations  is performed, it is not necessarily clear
			that  the mean, marginal variance or higher order moments are constant, even if there is  no change in the correlation. This paper develops change point analysis  for  the correlation structures under less restrictive assumptions.
			In contrast to previous work,   our approach does not require that the mean, variance and fourth order joint cumulants are constant under the null hypothesis. Moreover, we also address the problem of detecting relevant change points.
		}
		
		\bigskip
	\
		
	\end{abstract}
	



\section{Introduction}
\def\theequation{1.\arabic{equation}}
\setcounter{equation}{0}

Change point analysis  is a well studied subject in the statistical and econometric literature.  Since the seminal
work  on detecting structural breaks in the   mean of  \cite{page1954}  a powerful methodology has been developed to detect
various types  of change points in time series (see for example
\cite{auehor2013} and \cite{jandhyala2013}  for recent reviews of the literature).   Several authors have  argued that, in applications,
besides the mean  the detection of
changes in the variance or the correlation structure
of a time series  is of importance. Typical examples include the discrimination between stages of high and low asset
volatility or the detection of changes in the parameters of an AR($p$) model in order to obtain superior forecasting procedures.
\cite{wichern1976}  studied the change point problem for the variance in a first order
 autoregressive model. These authors pointed out that  - even if  log-return data exhibits a  stationary behavior in the mean -
 the variability is often not constant and
 as a consequence  any conclusions based on the assumption of homoscedasticity could be misleading.
\cite{abrwei1984} and \cite{bauras1985}  used  a Bayesian
 and an ML approach to find change points  in AR-models.  \cite{incltiao1994} proposed a nonparametric CUSUM-type test for changes in the variance of an independent
identically distributed sequence and \cite{leepark2001} derived corresponding results applicable to linear processes (see also
\cite{changevariance} who used the Schwarz information criterion. \cite{galpen2007} and
 \cite{aueetal2009}  suggested  nonparametric tests for structural breaks in the variance matrix of a multivariate time series.
There exist also several papers discusssing change point analysis in the second order structure of a time series.
For example,  \cite{bergomhor2009} and \cite{kileckjon2013} considered the more classical problem
of a change point in a correlation  at fixed  lag. 
 \cite{davis2006}  and  \cite{prepucdet2015} proposed methods for detecting multiple breaks in piecewise stationary processes. 
 
This list of references is by no means complete but an important and common feature of  the cited references and  most of the literature
on testing for structural breaks in the covariance or correlation structure (at different lags)
consists in the fact that the model is formulated such that the stochastic process under the null hypothesis of  ``no change-point''
is stationary. This assumption is crucial   to derive (asymptotic)  critical values for the corresponding testing procedures  using strong approximations or invariance principles. On the other hand this assumption drastically
restricts the applicability of the  methodology. For example,  \cite{incltiao1994}  and  \cite{aueetal2009}  assume for the construction of a testing procedure
for the hypothesis for change point in the variance
 that the mean of the sequence under consideration does not change in time (as the variance under the null hypothesis).
A similar assumption was made by
 \cite{wiekradeh2012}  in the context of testing for a constant  correlation, where the authors suggested
a CUSUM-type statistic for a change in the correlation of a stationary time series
if  at  the same time the means and variances do not change.
However, from a practical point of view,  assumptions of this type are very restrictive and there might be
many situations  where one is interested in a change of the  correlation  even if the mean
 and  the variances change gradually in time. In this case
the classical approach is not applicable.

The present paper is devoted to the construction of change point tests for the second-order characteristics of a non-stationary time series, in particular  changes in  the lag-$k$ correlation.
In Section \ref{sec2}  we introduce   piecewise locally stationary processes as considered  by \cite{zhou2013} who  investigated the properties of the classical CUSUM test for the mean under non-stationarity.
 Section \ref{sec3} is devoted to
 the ``classical'' change point problem  for  a (vector) of correlations at different lags   in a piecewise locally stationary process.
 In the simplest case of one lag-$1$ autocorrelation, say {\color{black}$\rho_i= {\emph Corr} (X_i,X_{i+1}) $} the hypothesis can be formulated as
 \begin{eqnarray} \label{nullvar}
H_0:\  \rho_i= \rho_j ~\ \text{ for all}\  i, j=1,\ldots , n  ~\text{ versus }\  H_1:\  \rho_i \neq \rho_j\ \text{for some}\ i\neq j.
\end{eqnarray}
 We propose a CUSUM approach based on nonparametric residuals and prove weak convergence of the corresponding CUSUM statistic. It turns out that the limiting distribution depends in a complicated  way on the dependence structure of the piecewise locally  stationary process, and for this reason a wild  bootstrap approach is developed and its consistency is
proved. The methodology is very general and applicable in many situations where the assumptions of classical tests are not satisfied.
In particular, we neither assume that the mean, variance or higher order joint cumulants of the non-stationary
sequence are constant nor that the change in the variance and the lag-$k$ correlation occur at the same location.
Furthermore, we show that the stochastic errors produced in the nonparametric estimation of the mean and
variance function are asymptotically negligible in the second-order CUSUM statistic. This result is of particular interest, and non-trivial because the order of the stochastic errors of the nonparametric estimates is larger than the $1/\sqrt{n}$ convergence rate of the
 CUSUM test.

The situation is more complicated if one is interested in such sophisticated hypotheses as
 {\it precise hypotheses} (see \cite{bergdela1987}).
Here  (in the simplest case)  one  assumes
  the existence of a change point  $k \in \{1,\ldots,n \}$
such that
\begin{align}\label{eq137a}
v_1=\rho_{1}=\ldots =\rho_{k}~\not = ~  v_2=\rho_{k+1}=\ldots  =\rho_{n}~,
\end{align}
and is interested in hypotheses   of the form
\begin{align}\label{relvar}
H_0: \Delta:=|v_2-v_1|\leq \delta\ \ \text{versus}\ \ H_1: \Delta:=|v_2-v_1|>\delta
\end{align}
for  some pre-specified constant $\delta>0$. Throughout this paper we call hypotheses of the form
\eqref{nullvar} {\it ``classical''} in order to distinguish these from the precise hypotheses of the form \eqref{relvar}. Although hypotheses of the form \eqref{relvar} have been discussed in other fields (see \cite{chowliu1992} and \cite{mcbride1999})
the problem of testing precise hypotheses has only recently been considered  by
\cite{dettwied2016} in the context of change point analysis.  These authors point  out that in many cases  a modification
of the  statistical analysis might not be necessary if a change point has been identified but the
difference between the parameters
before and after the change-point is rather small.
In particular, inference might be robust under ``small'' changes of the parameters  and
changing decisions (such as trading strategies or modifying a manufacturing process) are
expensive and should therefore only be performed if changes have serious consequences.
 Testing a hypothesis  of the form \eqref{relvar} to detect a structural break
  also avoids the consistency problem mentioned in \cite{berkson1938}: any test will detect negligible
 changes in the parameter if the sample size is sufficiently large.
 \cite{dettwied2016} call the hypotheses of the form \eqref{relvar} hypotheses of a  {\it non-relevant} (null hypothesis)  and
 {\it relevant change point} (alternative)
 and, according to their argumentation, only  relevant change points  should be detected, because one has to distinguish scientific from statistical significance.
   
 Although the testing problem in the form \eqref{relvar}
 is appealing, the construction of corresponding tests faces several   mathematical challenges. In particular,
  even under the null hypothesis of a {\it non-relevant change point}
  one has to deal with the problem of non-stationarity.
For example,
 \cite{dettwied2016}    developed a CUSUM-type test for    the hypotheses in \eqref{relvar} which is only  applicable
 under the assumption that the time series before and after the
change point is strictly stationary. In the context of change point analysis for correlations this means that the mean and the variances
of the process have to be constant before and after the change point.
From a practical point of view this assumption
seems to be very strong and not very realistic.

Section \ref{sec4} is devoted to the problem of testing the hypothesis of a non-relevant change in the several
correlations at different lags. We use the CUSUM approach proposed in \cite{dettwied2016} to obtain a test for the hypothesis \eqref{relvar} and its analogue in the case of lag-$k$ correlations. Asymptotic normality of a corresponding $\mathcal{L}_2$-type statistic is established and   a wild bootstrap method is developed that addresses the particular structure of the hypotheses in relevant change point analysis.
 To  our best knowledge  resampling procedures  for this type of change point analysis in non-stationary nonparametric problems have not
    been considered in the literature.
 The finite sample properties of the new procedures are investigated by means of a simulation study in Section \ref{sec5}. In Section \ref{sec:data} we analyze the USD/CAD exchange rate series and illustrate the usefulness of the proposed methodology in identifying second order change points in modeling  volatilities.  All proofs and technical details are deferred to an online supplement (see also  \cite{holger2015}).

\section{Piecewise locally stationary processes} \label{sec2}
\def\theequation{2.\arabic{equation}}
\setcounter{equation}{0}

We start writing some notations of frequent use.
For an $l$-dimensional (random) vector $\mathbf v = (v_1,...,v_l)$, $l\geq 1$, let $|\mathbf v| =(\sum_{i=1}^lv_i^2)^{1/2}$.
A random vector $\mathbf V$ is said to be in $\mathcal L_q$ , $q > 0$, if $\E(|\mathbf V |^q ) < \infty$. In this case
write $\|\mathbf V\|_q=(\E|\mathbf V|^q)^{1/q}$, and $\|\mathbf V\|=\|\mathbf V\|_2.$
The symbol   $\cod $  means weak convergence of real-valued random variables (convergence in distribution).
 For any interval  $\mathcal{I} \subset \R $ and nonnegative integer $q$ let $ \mathcal{C}^{q}( \mathcal{I})$
 be the set of $q$  times continuously differentiable
functions $f:  \mathcal{I} \to \R $ and $\mathcal{C} (\mathcal{I})=\mathcal{C}^0(\mathcal{I})$.
Let $\{\varepsilon_i\}_{i \in \mathbb{Z}}$ denote a sequence of independent identically distributed (i.i.d.) random
variables and  denote by $\FF_i=\sigma(...,\varepsilon_0,...,\varepsilon_{i-1}\varepsilon_i)$ the sigma field generated by
$\{ \varepsilon_j | j \leq i  \}$. We define the sigma field
$\FF_i^{(j)}=\sigma(...,\varepsilon_{j-1}\varepsilon'_{j}\varepsilon_{j+1}...,\varepsilon_i)$, where $\{\varepsilon'_i\}_{i \in \mathbb{Z}}$ is an independent copy of
$\{\varepsilon_i\}_{i \in \mathbb{Z}}$, and $\FF_i^*=\FF_i^{(0)}$ for short.
 For any real number $a$, write $\lf a\rf$ be the largest integer which $\leq a$. Let $\mathbf 1(\cdot)$ be the indicator function, $sign(\cdot)$ be the usual sign function, such that $sign(x)=\mathbf 1(x\geq 0)-\mathbf 1(x<0)$. Define $0/0=1$.
Let $a\wedge b$ denote $\min(a,b)$ for $a,b\in \mathbb R$. Through out the paper we consider the case that type I error $\alpha\leq 0.05$. {\color{black}We discuss autocorrelation in the rest of the paper, and use the term ``correlation'' for ``autocorrelation'' for short. Our method can be applied to cross correlation without further difficulty.}

We consider the model
\begin{align}\label{mod1}
Y_i=\mu(t_i)+e_i, \qquad i=1,\dots,n,
 \end{align}
 where (for the sake of simplicity) $t_i=i/n$ ($i=1,\ldots ,n$) and $\mu(\cdot)$ is a smooth function.
Formally $\{Y_i\}_{i=1}^n$ is a triangular array of random variables but  we do not reflect this fact in our notation.
 Change point problems for this  model have found considerable attention in the recent literature,
 where most of the work refers to problems of detecting changes of the mean in the situation of centered  and independent identically distributed
 (i.i.d.)  errors (even assumed to be Gaussian in some cases)
 (see \cite{Mueller1992} for an early reference and    \cite{Mallik2011} and \cite{Mallik2013}  for more recent references). 
 \cite{vogtdett2015} proposed  a generalized CUSUM approach to detect  gradual changes in model \eqref{mod1} using a different concept of local stationarity (see \cite{vogt2012}).
  \\
 In the present  paper we consider non-stationary processes of the form
 \eqref{mod1}
 and  are interested in identifying abrupt changes in
 the correlations.
 \ More precisely we consider
 an error process $\{e_i\}_{i=1}^n$  in   \eqref{mod1} that is
piecewise locally stationary ($PLS$)  with $r$ breaks for some $r \in \mathbb{N}$. Formally, we use a definition for a $PLS$ process
and  the concept of ``\emph{physical dependence measure for PLS}'' that is given in \cite{zhou2013}.
\begin{definition} \label{def1} ~~\\
(1) The sequence $\{e_i\}_{i=1}^n$ is called PLS with $r$ break points  if there exist constants $0=b_0<b_1<...<b_r<b_{r+1}=1$ and nonlinear filters $G_0, G_1,..., G_r$, such that
{\color{black}\begin{align*}
e_i=e_i(t_i), \text{where}\  e_i(t)=G_j(t,\FF_i),  \mbox{ if } b_j<t_i\leq b_{j+1}
\end{align*}}
where  $\FF_i=\sigma(...,\varepsilon_0,...,\varepsilon_{i-1} \varepsilon_i)$, and $\{\varepsilon_i\}_{i \in \mathbb{Z}}$ is a sequence of  i.i.d. random variables.

(2) Assume that $\max_{1\leq i\leq n}\|e_i\|_p<\infty$ for some  $p \ge 1 $.
 Then for $k>0$, define the $k_{th}$ physical dependence measure in $\mathcal{L}_p$-norm as
\begin{align*}
\delta_p(k)=\max_{0\leq i\leq r}\sup_{b_i< t\leq b_{i+1}}\|G_i(t,\FF_k)-G_i(t,\FF_k^*)\|_p,
\end{align*}
where   $\delta_p(k)=0$ if $k<0$.

\end{definition}

The $PLS$ process is a natural non-stationary extension of many well known statistical processes, with the dependence measure easy to calculate. 
\begin{example}(PLS linear process)  \rm
For $\{\varepsilon_i\}_{i\in \mathbb Z}$ take $\FF_i=\sigma(\{\varepsilon_j|j\leq i\})$, and consider the
 process
\begin{align}\label{PLSlinear}
G_j(t,\FF_i)=\sum_{s=0}^\infty a_{j,s}(t)\varepsilon_{i-s}\ b_{j}< t\leq b_{j+1}  \ 0\leq j\leq r,
\end{align}
where $0=b_0< b_1< ...< b_{r+1}=1$ are unknown break points, $a_{j,s}(t)$ for $b_{j}< t\leq b_{j+1}  \ 0\leq j\leq r$, $s\in \mathbb Z$ are Lipchitz continuous functions. Straightforward calculations show that  $\delta_p(k)=O(\max_{0\leq j\leq r}\sup_{b_j< t\leq b_{j+1}}|a_{j,k}(t)|)$ provided $\|\varepsilon_0\|_p<\infty$. Model \eqref{PLSlinear} is a time-varying MA process with possible abrupt changes. For smooth time-varying MA process, it could be shown, for example in \cite{zhang2012} that it well-approximates the locally stationary autoregressive processes that
have been studied extensively in the literature (see for example  
\cite{Dahlhaus1997} among others).
\end{example}
\begin{example}(PLS nonlinear process) \rm
For $\{\varepsilon_i\}_{i\in \mathbb Z}$ take  $\FF_i=\sigma(\{\varepsilon_j|j\leq i\})$ and consider the  process
\begin{align}\label{PLSnonlinear}
G_j(t,\FF_i)=R_j(t,G_j(t,\FF_{i-1}),\varepsilon_i),\ b_{j}< t\leq b_{j+1}  \ 0\leq j\leq r,
\end{align}
where $0=b_0< b_1< ...< b_{r+1}=1$ are unknown break points. Many important nonlinear time series have the form $X_i=R(X_{i-1}\varepsilon_i)$.
Typical examples include  (G)ARCH models 
(see \cite{Engle1982} \cite{Bollerslev1986}), threshold models
(see \cite{Tong1990}) and bilinear models. It can be shown similarly to
\cite{zhouwu2009} that, under some mild conditions, $\delta_p(k)=O(\chi^k)$ for some $\chi\in (0,1)$, and that $\chi$ can be evaluated as
\begin{align}
\chi:=\max_{0\leq j\leq r}\sup_{t\in (b_j,b_{j+1}]} \sup_{x\neq y}\frac{\|R_j(t,x,\varepsilon_0)-R_j(t,y,\varepsilon_0)\|_p}{|x-y|}.
\end{align}
\end{example}

For our asymptotic analysis we list some conditions.
\begin{description}
\item (A1) The process $\{e_i\}^n_{i=1}$ is $PLS$ and {\it piecewise stochastic Lipschitz continuous} with $r$ break points: there exists a constant
$C > 0$ such that, for all $i\in\{0,\dots,r\}$  and all $t,s\in (b_i,b_{i+1}]$,  \begin{align*}
\|G_i(t,\FF_0)-G_i(s,\FF_0)\|_\iota\leq C |t-s|
  \end{align*}
  holds for   $\iota\geq 8$. 
   In addition, $\E[e_i]=0$ for all $1\leq i\leq n$, and there is a  \emph{variance function} $\sigma^2(\cdot): [0,1] \to \mathbb{R}^+$, such that $\sigma^2_i := \sigma^2(t_i)= {\rm Var}(e_i)$, for $i=1,\dots,n.$
\item (A2) The second derivative $\ddot{\mu}(\cdot)$ of the function $\mu(\cdot)$ in  model \eqref{mod1}  exists and is Lipschitz continuous on the interval $[0,1]$.
\item (A3) $\max_{0\leq i\leq r}\sup_{t\in(b_i,b_{i+1}]}\|G_i(t,\FF_0)\|_\iota<\infty$ for some $\iota\geq 8$. 
\item (A4) $\delta_\iota(k)=O(\chi^k)$ for some $\chi\in (0,1)$ and some $\iota\geq 8$.
\end{description}
\begin{remark} {\rm ~~\\
a)  The bound of $\max_{1 \leq i \leq n}\|e_i\|_p$ in Definition \ref{def1} does not depend on $n$. This
 simplifies the assumptions and the  proofs in the subsequent discussion. It is also possible to develop corresponding results for
an $n$-dependent bound with added complications
in the technical arguments. \\
{b)
For the sake of brevity we  use the condition  $\iota\geq 8$ in (A3) and (A4). Using additional technical arguments
it can be shown that
our methodology is still
valid for innovations with a  heavier tail (see also Section \ref{sec5} for some simulation results with heavy-tailed distributions).}\\
c)
The process  $\{e_i^2\}^n_{i=1}$  of squared errors is also $PLS$. Simple calculations show  that $\{e_i^2\}^n_{i=1}$ satisfies the assumptions (A1), (A3), (A4) with $\iota\geq 4$.
}
\end{remark}
\section{Tests for changes in correlations } \label{sec3}
\def\theequation{3.\arabic{equation}}
\setcounter{equation}{0}

Suppose that we observe data $\{Y_i\}_{i=1}^n$ according to model \eqref{mod1} where the process $\{e_i\}_{i=1}^n$ is $PLS$  and $\mu(\cdot)$ is
an unknown deterministic trend.
We are interested in testing nonparametrically the ``classical'' hypothesis of a  change point in
 the correlations.
The important difference to previous
work on this subject (see for example \cite{incltiao1994} or \cite{aueetal2009}) is that in  general the process is NOT assumed to be stationary under the
 null hypothesis of no  change point.  This means - for example - that the approach proposed here can be used to test the hypotheses \eqref{nullvar} where the mean is not constant. The price for this type of flexibility is that critical values of the asymptotic distribution of the CUSUM statistic are not directly available.
For this we develop  a bootstrap CUSUM-type  test for the ``classical'' hypotheses of a change point in correlations based on residuals from a local linear fit.  For the definition of the local
linear estimator we assume  that the corresponding kernel function, say $K$,
is  symmetric  with support $[-1,1]$ satisfying  $\int K(x)dx=1$,
 and define for $b>0$ the function  $K_{b}(\cdot)=K(\frac{\cdot}{b})$.
  We assume that  $K \in \mathcal{C}^{2}([-1,1])$. 
 and, for convenience, we set $e_i=0$, $\hat e_i=0$ if $i>n$, where $n$ is the sample size.

Consider the problem of testing whether there are changes in correlations $\rho_{i,k}:=Corr(Y_i,Y_{i+k})$ for some pre-specified lag-$k$'s, with
 \begin{align}
&H_0:\  \rho_{i,k}=\rho_{j,k}=\rho_k\ \text{ for all}\  i, j=1,\ldots , n, k=r_1,\dots,r_l ~ \label{mulcorr0}  \\
 & H_1: \text{  There  exists  } 1\leq s\leq l \text{ and } i\neq  j  \text{  such that }  \ \rho_{i,r_s}\neq \rho_{j,r_s}\   \label{mulcorr1}
\end{align}
where the integers $r_1<r_2<\cdots<r_l$ define the lags of interest.
  A test for the classical hypothesis 
  for  stationary processes can be derived by similar arguments as given
  in \cite{wiekradeh2012} under the additional assumption that the mean and variance  are not changing. However,
 statistical  inference regarding changes the correlation structure  in a  locally stationary framework  (including non constant mean or variance)
 requires estimates of the mean and variances.
For this purpose  
 consider the CUSUM statistic
\begin{align}\label{hatTn}
\hat{T}_n=\max_{1\leq i\leq n}\Big|\hat{\mathbf S}^{}_i-\frac{i }{n}\hat{\mathbf S}^{}_{n}\Big|,
\end{align}
where $\hat{\mathbf S}_i=(\hat S^{(r_1)}_i,...,\hat S^{(r_l)}_i)$, $\hat{S}^{(j)}_i=\sum_{s= 1}^i\frac{\hat{e}_s\hat{e}_{s+j}}{\hat \sigma^2(t_s)}$, 
$\hat{e}_s= Y_s-\hat{\mu}_{b_n}(t_s)$, $\hat{e}_{s+j}= Y_{s+j}-\hat{\mu}_{b_n}(t_{s+j})$, and $\hat{\mu}_{b_n}({\cdot})$ is the local linear estimator of
the function  $\mu(\cdot)$ with bandwidth $b_n$, 
 \begin{align}\label{defhatmu}
(\hat{\mu}_{b_n}(t), \hat{\dot{\mu}}_{b_n}(t))=\argmin_{\beta_0,\beta_1} \sum^n_{i=1}\big(Y_i-\beta_0-\beta_1(t_i-t)\big)^2K_{b_n}(t_i-t)
\end{align}
(see \cite{fangij1996}). 

We allow the variance to possibly have a
 structural break at a point, say  $\tilde{t}_v$ that need not coincide with the location of the change point in any of the lag-$k$ correlations. We assume that $\ddot \sigma^2(\cdot)$ is Lipschitz continuous
on the intervals  $(0,\tilde{t}_v)$ and $(\tilde{t}_v,1)$     and that there exists a constant $\zeta>0$, such that $\tilde{t}_v\in[\zeta,1-\zeta]$.
We define    an  estimator, say $t_n^* $, of the change point $\tilde{t}_v$ in the variance by
\begin{align} \label{tstar}
 t^*_n=\argmax_{\lf {n\zeta}\rf\leq i\leq n-\lf n\zeta\rf+1}|\mathcal{M} (i)|/n,
\end{align} where
\begin{align} \label{mi}
\mathcal{M}(i)= \frac {1}{L}\Big(\sum_{j=i-L+1}^i\hat e_j^2-\sum_{j=i}^{i+L-1}\hat e_j^2\Big)
\end{align}
and
$L \in \mathbb{N} $ is a regularization parameter that increases with $n$. The maximum in \eqref{tstar} is not taken over the full range $1 \leq i \leq n$, as recommended in \cite{andrews1993}  (see also \cite{qu2008}).
{
We estimate $\sigma^2(t_i)$ by
$
\hat{\sigma}^2(t_i)=\hat{\sigma}^2_{c_n,b_n}(t_i,n t_n^*),
$
where for $k=1,\ldots ,n$
 $$
 \hat{\sigma}^2_{c_n,b_n}(t,k)=\hat{\sigma}^2_{c_n,b_n}(t,k-)\mathbf 1(t\leq k/n)+\hat{\sigma}^2_{c_n,b_n}(t,k+)\mathbf 1(t>k/n)
 $$
 and
\begin{align}\label{defhatmu1}
(\hat{\sigma}^2_{c_n,b_n}(t,k-),\widehat{\dot{\sigma^2}}_{c_n,b_n}(t,k-))=\argmin_{\beta_0,\beta_1}\sum_{i=1}^{k}(\hat{e}^2_{i}-\beta_0-\beta_1(t_i-t))^2K_{c_n}(t_i-t),\notag\\
(\hat{\sigma}^2_{c_n,b_n}(t,k+),\widehat{\dot{\sigma^2}}_{c_n,b_n}(t,k+))=\argmin_{\beta_0,\beta_1}\sum_{i={k+1}}^{n}(\hat{e}^2_{i}-\beta_0-\beta_1(t_i-t))^2K_{c_n}(t_i-t).
\end{align}
We take the  (non-observable) analogue of $ \hat S_i^{(j)}$ to be
 \begin{align}\label{70a}
   S_i^{(j)}=\sum_{s=1}^iW_s^{(j)}~,
\end{align}
where $W_s^{(j)} = \frac {e_se_{s+j}}{\sigma(t_s)\sigma(t_{s+j})} $, and consider the random variable
\begin{align}\label{Tn}
{T}_n=\max_{1\leq i\leq n-r_l}\Big| {\mathbf S}^{}_i-\frac{i }{n} {\mathbf S}^{}_{n}\Big|,
\end{align}
where ${\mathbf S}_i=( S^{(r_1)}_i,...,S^{(r_l)}_i)$.
}
It is  easy to see that $W_i^{(j)}$   is $\FF_{i+j}$ measurable and that the process $(W_i^{(j)})^{n-j}_{i=1}$ is $PLS$. Moreover, $\{W_i^{(k)} k=r_1,...,r_l\}_{1\leq i\leq n-r_l}$ can be modeled by an $l$-dimensional $PLS$ process.
Take $q$ as the number of break points,  $0=v_0<v_1<...<v_{q+1}=1$ as the corresponding locations of the breaks, and ${\mathbf H}$
as the corresponding nonlinear filters, $(W_i^{(r_1)}...W_i^{(r_l)})^T = {\mathbf H}_j(t_i, \FF_{i+r_l})$ if $v_j < t_i \leq v_{j+1}$, $0\leq j\leq q$.

 The following result shows that $t_n^*$ is a consistent estimate of $\tilde{t}_v$;
 its proof can be found in Section \ref{sec61} of the online supplement.
 \begin{lemma}\label{tstar0}
  Assume that $nb^6_n\rightarrow 0$, $nb^3_n\rightarrow \infty$ and  that
 (A1) - (A4) are satisfied with $\iota>8$. Suppose that the variance function is
twice differentiable on the intervals $(0,\tilde{t}_v)$ and $(\tilde{t}_v,1)$, such that the second derivative $\ddot{\sigma^2}(\cdot)$
is Lipschitz continuous  (here $\tilde{t}_v$ is the location of the change point of the variance such that
$\zeta\leq \tilde t_v\leq 1-\zeta$).
Then  the estimator  $t_n^*$ defined in \eqref{tstar} satisfies
$t^*_n-\tilde{t}_v=o_p(n^{-(1-4/\iota)}\log n)$.
 \end{lemma}
  \begin{remark}{\rm 
  The  rate of convergence of the estimator $t_n^*$ is arbitrarily close  to the optimal rate $n^{-1}$ subject to a logarithmic factor if (A1) and (A4) hold for any $\iota > 0$.  }
 \end{remark}

{The  rates of  convergence of the estimators  \eqref{defhatmu} and \eqref{defhatmu1} are of the order $n^{-2/5}$
under suitable bandwidth conditions. Thus, a naive plug-in argument of $\hat \mu(t_i)$ does not lead to the crucial result that
 \begin{equation}
 \label{approx1}
 |\hat T_n-T_n|=o_p(\sqrt n),
 \end{equation}
 that is required for constructing the hypothesis testing procedure.
In the Appendix we demonstrate that  the estimate  \eqref{approx1} is in fact valid using
 delicate arguments to overcome the slow rate of  convergence of the non-parametric fit.
Then the weak convergence of the statistic $\hat T_n/\sqrt{n}$   follows  from the weak convergence of $ T_n/\sqrt{n}$, which can be
established under an additional assumption.
\begin{description}\item (A5) The long run variance function
 \begin{equation} \label{sigma1}
\kappa^2(t)=\sum_{k=-\infty}^\infty \text{cov}(\mathbf H_i(t,\FF_k),\mathbf H_i(t,\FF_0))\in \mathbb R^{l\times l} \qquad \text{if $t\in(v_i,v_{i+1}]$}\ 0\leq i\leq q,
\end{equation}
and $\kappa^2(0):=\lim _{t\downarrow0} \lambda_{\min}(\kappa^2(t))$ exists with $\inf_{t\in[0,1]}\lambda_{\min}(\kappa^2(t))>0$, where for any positive semi-definite matrix $A$, $\lambda_{\min}(A)$ denotes the minimal eigenvalue of matrix $A$.
\end{description}
The  proof
of the following result is deferred to the online appendix.

 \begin{theorem}\label{theorem1}
Assume that {$b_n\rightarrow 0$, $c_n/b_n\rightarrow 0$, $c_nb_n^{-2}\rightarrow \infty$, $nc_n^4\rightarrow 0$, $nb_n^{6}c_n^{-1/2}\rightarrow 0$, $nb_n^{4}c_n^{1/2}\rightarrow \infty$} and suppose that (A1) - (A5) are satisfied with $\iota\geq 8$.
Assume that the variance $\sigma^2(t)>0$. Suppose that one of the  the following conditions is satisfied.
\begin{description}
\item(i)  $\sigma^2(\cdot)$ is twice differentiable on [0,1] and the second derivative  $\ddot{\sigma}^2(\cdot)$ is Lipschitz continuous.
\item(ii) $\sigma^2(\cdot)$ has one abrupt change point $\tilde t_v\in [\zeta,1-\zeta]$,
 and on the intervals $[0,\tilde{t}_v)$ and $(\tilde{t}_v,1]$, $\sigma^2(t)$ is twice differentiable and the second derivative $\ddot{\sigma}^2(\cdot)$
is Lipschitz continuous. \end{description}
Then
 under the null hypothesis \eqref{mulcorr0} we have
\begin{align}\label{eqthm1}
\frac{1}{\sqrt{n}}\hat T_n \cod  \mathcal{K}_1  := \sup_{t\in(0,1)}| \mathbf U(t)-t\mathbf U(1)|,
\end{align}
where  $\{\mathbf U(t)\}_{t \in [0,1]}$  is a zero mean $l-$dimensional Gaussian process with covariance function
\begin{equation}\label{cov1}
\gamma(t,s)=\int_0^{\min(t,s)}\kappa^2(r)dr.
\end{equation}
\end{theorem}

\medskip

As a consequence of Theorem \ref{theorem1} we obtain - in principle - an asymptotic level $\alpha$ test for the hypothesis \eqref{nullvar} by rejecting $H_0$, whenever
$
\frac {1}{\sqrt{n}} \hat T_n > q_{1- \alpha}
$
where $q_{1 - \alpha}$ is the $(1- \alpha)$-quantile of the distribution of the random variable $\mathcal{K}_1$ in \eqref{eqthm1}.
However, under non-stationarity (more precisely under the $PLS$ assumption),
the function $\kappa^2(t)$ defined in \eqref{sigma1} and, as a consequence, the covariance structure of the Gaussian process  $ \{  \mathbf U(t)-t \mathbf U(1)\}_{t\in[0,1]} $
involves the complicated dependence structure of the data generating process.

Due to the $PLS$ structure, the covariance structure of the
Gaussian process $\mathbf U(\cdot)$ and the quantiles of the limiting distribution in Theorem  \ref{theorem1}
are  hard to estimate.
As an alternative,  a data-driven critical value will be derived using  a wild bootstrap method to mimic the distributional properties of the Gaussian process $\mathbf U(\cdot)$.
Following   \cite{zhou2013} we
define for  a fixed window size, say $m$,   the quantities
\begin{align}\label{new.83}
\hat {\mathbf \Phi}_{i,m}=\frac{1}{\sqrt{m(n-m+1)}}\sum_{j=1}^i\Big(\hat {\mathbf S}_{j,m}-\frac{m}{n}\hat{\mathbf S}_{n}\Big)R_j, \qquad %
i=1,...,n-m+1,
\end{align}
where $\hat{\mathbf S}_{j,m}=(S^{(r_1)}_{j,m}...,S^{(r_l)}_{j,m})^T$, $\hat{\mathbf S}_{n}=\hat{\mathbf S}_{1,n}$, $\hat S^{(k)}_{j,m}=\sum_{r=j}^{j+m-1}\frac{\hat e_r\hat e_{r+k}}{\hat \sigma^2(t_r)}$, and $\{R_i\}_{i\in \mathbb{Z}}$ is a sequence of i.i.d
standard normal distributed random variables independent of $\{ \varepsilon_i \}_{i \in \mathbb{Z}}$.
\begin{theorem}\label{theorem2}If the conditions of Theorem \ref{theorem1}   are satisfied and, for $m\rightarrow \infty$, assume $m/\sqrt n\rightarrow 0$,  { $\sqrt{m}\big(c_n^2+(\frac{1}{\sqrt{nc_n}}+b_n^2+\frac{1}{\sqrt{nb_n}})c_n^{-1/4}\big)\log n\rightarrow 0$}
(conditional on $\FF_n$ in probability)
\begin{align*}
M_n= \max_{m+1\leq i\leq n-m+1}\Big|{\hat {\mathbf \Phi}}_{i,m}-\frac{i}{n-m+1}{\hat {\mathbf \Phi}}_{n-m+1,m}\Big |\cod  \mathcal{K}_1,
\end{align*}
where the random variable $\mathcal{K}_1$ is defined in \eqref{eqthm1}.
\end{theorem}
\noindent
Theorem \ref{theorem2}  provides an asymptotic level $\alpha$ test for the hypothesis of constant correlations in model \eqref{mod1} with critical values obtained by resampling. 
The proof
is deferred to the online supplement.  The details of generating the critical values and performing the test are summarized in an algorithm.
\begin{algorithm} \label{algvar}
~~\\{\rm
[1] Calculate the statistic $\hat{T}_n$ at (\ref{hatTn}).

\noindent
[2] Generate $B$ conditionally $i.i.d$ copies $\{\hat{\mathbf \Phi}^{(r)}_{i,m}\}_{i=1}^{n-m+1}$ $(r=1,\dots,B)$ of the random variables
$\{\hat{\mathbf \Phi}_{i,m}\}_{i=1}^{n-m+1}$  defined in \eqref{new.83} and calculate
 $$M_r=\max_{m+1\leq i\leq n-m+1}\Big|\hat{\mathbf \Phi}_{i,m}^{(r)}-\frac{i}{n-m+1}\hat{\mathbf \Phi}^{(r)}_{n-m+1,m}\Big|.$$
[3] If $M_{(1)}\leq M_{(2)}\leq ... \leq M_{(B)}$ denote the order statistics of   $M_1,\dots,M_B$, null hypothesis
of constant correlations is rejected   at level $\alpha$ when
 \begin{equation} \label{testvar}
    \hat{T}_n/\sqrt{n}>M_{\lf B(1-\alpha)\rf}.
    \end{equation}
 The $p$-value of this test  is given   by $1-\frac{B^*}{B}$, where $B^*=\max\{r: M_{(r)}\leq \hat{T}_n/\sqrt{n}\}$.
 }
\end{algorithm}

\begin{remark}\label{consis_test}  {\rm ~\\
 {\color{black} (1) If  the sequence $b_n$ is of order $n^{-1/5}$ and $m$ is of  order $n^{1/3}$, then the bandwidth conditions of Theorem \ref{theorem2} hold
if   the sequence $c_n$ is of order $n^{-\beta}$, where $\beta \in ( \frac{1}{4}\frac{2}{5})$.}  \\
(2)
It follows by similar arguments as in the proof of Theorem 2, Proposition 3 of \cite{zhou2013} and Lemma \ref{mu} and   Lemma \ref{boundhat}
 in the online supplement,   that the bootstrap test \eqref{testvar} is consistent. For $1\leq s\leq l$, write $\rho_{r_s}(t_i)=\rho_{i,r_s}$, $\mathbf \rho(\cdot)=(\rho_{r_1}(\cdot),...\rho_{r_l}(\cdot))^T$. It can be shown that the  bootstrap is able to detect  local alternatives
  of the form $\rho(\cdot) = \rho_0+n^{-1/2}f(\cdot)$, where $f(\cdot)$ is a nonconstant piecewise Lipschitz continuous $l-$dimensional vector function.\

  }
\end{remark}

\section{Relevant changes of correlations} \label{sec4}
\def\theequation{4.\arabic{equation}}
\setcounter{equation}{0}

After a change point has been detected and localized a modification of the statistical analysis  is necessary, one that addresses
the different features of the data generating process before and after the change point. \cite{dettwied2016} pointed out that,
in many cases, such a modification might not be necessary if the difference between the parameters
before and after the change point is rather small. Inference might be robust with respect to small changes of the correlation structure,
but changing decisions (such as trading strategies or modifying a manufacturing process) might be very expensive and only be performed if changes would have serious consequences.
 Here we investigate  the hypothesis \eqref{relvar} of a {\it non-relevant change point} 
 for correlations 
 in a general non-stationary context  under the assumption  of PLS.

 Consider model \eqref{mod1} and suppose that
 there exist  time points $t_k\in (0,1)$, $k=r_1,...,r_l$,   such that
\begin{align*}
\rho_1^{(k)}=\rho_{1,k}= ... =\rho_{\lf nt_k\rf,k}\quad \quad \rho_2^{(k)}=\rho_{\lf nt_k\rf+1,k}= ... =\rho_{n,k}.
\end{align*}
We are interested in testing the hypotheses
\begin{align}\label{hypo3}
&H_0:\  |\rho_{1}^{(k)}-\rho_{2}^{(k)}|\leq \delta_k \text{ for all}\  k=r_1,\dots,r_l  ~ \\
& H_1: \text{There exists a lag }  k\in \{r_1,\ldots , r_l\} \text{ such that } \ |\rho_{1}^{(k)}-\rho_2^{(k)}|>\delta_k,
\end{align}
where $\delta_{r_1}\ldots , \delta_{r_l}$ are given thresholds. 
{Problems of this type  have recently been discussed in \cite{dettwied2016} under  assumptions that are not practically tenable. 
In  the  PLS framework, these assumptions will be relaxed. 
However, under these more general assumptions,  the construction of a test and the investigation of its asymptotic properties
  is  substantially more difficult, as described 
   in the following paragraphs.
}

We denote by, for $1\leq s\leq l$, ${\Delta}_{r_s}=\rho^{(r_s)}_2-\rho^{(r_s)}_1$  the (unknown) difference before and after the change point
and assume here that, under the null hypothesis of
a non-relevant change in the correlations, the variance function $\sigma^2(\cdot)$  has either no jumps or has a jump at a  point, say
$\tilde{t}_v$, that need not coincide with any of the change  point $t_k$  in the correlation structure.
We define the CUSUM process, for $k=r_1,...,r_l$, by
\begin{align} \label{Un}
\hat{\mathcal V}^{(k)}_n(s)=\frac{1}{n}\sum_{j=1}^{\lf ns\rf}\frac{\hat{e}_j\hat{e}_{j+k}}{  \hat{\sigma}^{2}(t_j)}-\frac{\lf ns \rf}{n}\sum_{j=1}^{n}\frac{\hat{e}_j\hat{e}_{j+k}}{  \hat{\sigma}^2(t_j)}
\end{align}
where $\hat e_i = Y_i - \hat \mu_{b_n}(t_i)$ denotes the nonparametric residuals from the local linear fit
while using the convention that $\hat{e}_i=0$ for $i> n$.
The estimator for the change point of the correlation structure at lag-$k$ is taken to be
\begin{equation} \label{changeestcorr}
\hat{t}^{(k)}_n=  \argmax_{1\leq m\leq n}\big (\hat {\mathcal V}_n^{(k)}(m/n) \big)^2/n.
\end{equation}
The statistic $\hat t^{(k)}_n$ depends on the estimator $ t_n^* $ for  the change point in the variance as defined in
\eqref{tstar}. 
The estimator is consistent (a proof can be found in the online supplement.)

\begin{lemma}\label{lemvar}
Suppose that one of the following conditions holds.
 \begin{description}\item(i) Conditions of Lemma \ref{tstar0} are satisfied.
 \item(ii) $\sigma^2(\cdot)$ is twice differentiable on $ [0,1] $  and the second derivative $\ddot{\sigma}^2(\cdot)$ is  Lipschitz continuous.
  \end{description}
   In addition, suppose the conditions for the bandwidths $b_n$ and $c_n$ of Theorem \ref{theorem1} hold. Then for any $k=r_s$, $1\leq s\leq l$,  
the estimate $\hat t^{(k)}_n$
of the change point in the correlation structure at lag-$k$ defined by \eqref{changeestcorr}
satisfies
\begin{align}
& \hat{t}_n^{(k)}  \cod T^{(k)}_{\max}\ { \text{if}\  | \Delta_k |=0}\label{new.116} \\
|\hat{t}^{(k)}_n-t_k| &  =O_p(n^{-\upsilon}),\ { \text{if}\ | \Delta_k |>0}\label{new.115}
\end{align}
for some $\upsilon \in ( 1/2,2/3) $, where $T^{(k)}_{\max}$ is a $[0,1]$-valued random variable.
\end{lemma}
The test for the hypothesis of a non-relevant change is based on the statistic   \begin{align}\label{76}
\hat{T}^{(k),r}_n=\frac{3}{(\hat{t}^{(k)}_n)^2(1-\hat{t}^{(k)}_n)^2}\int_{0}^{1} (\hat{\mathcal V}^{(k)}_n(s))^2 ds,
\end{align}
where the the process $\{\hat{\mathcal V}_n^{(k)}(s),0\leq s\leq 1\}$ is defined in \eqref{Un}.
We show that $\hat{T}^{(r_s),r}_n$ is a consistent estimator of  $\Delta_{r_s}^2 = (\rho_{1}^{(r_s)}-\rho_{2}^{(r_s)})^2$ for $s=1,\ldots ,l$, and provide
its  asymptotic  distribution.

\begin{theorem}\label{thm6}
Assume  {that the   conditions for the bandwidths $b_n$ and $c_n$ of Theorem \ref{theorem1} hold}  and  that (A1) - (A4) are satisfied with {$\iota\geq 16$}. 
\begin{description}
\item(i) If $\Delta_k \neq 0$ for $k=r_1,...,r_l$, then
  {\begin{align}\label{revision2-163}
& \Big \{ \sqrt{n} \ \frac{\hat {T}^{(r_s),r}_n-\Delta_{r_s}^2}{|\Delta_{r_s}|}\Big \} _{s=1}^l~\cod ~\mathcal{Z} :=
 \{\mathcal{Z}^{(r_s)}\frac{\Delta_{r_s}}{|\Delta_{r_s}|} \}_{s=1}^l~,
\end{align}}
where
 \begin{align}
  \label{revision2-163B}
& \mathcal{Z}^{(r_s)}:= \frac{6}{{t_{r_s}}^2(1-{t_{r_s}})^2}\int_0^1\left(\mathbf U^{(s)}(u)-u\mathbf U^{(s)}(1)\right)\left(ut_{r_s}-u\wedge t_{r_s}\right){} du,
\end{align}
with  the  process $\{\mathbf U(u)\}_{u \in [0,1]}  =  \{ (\mathbf U^{(1)}(u) , \ldots , \mathbf U^{(l)}(u))^T \}_{u \in [0,1]} $
as defined in Theorem \ref{theorem1}.
\item(ii) If $\Delta_{r_s} = 0$ for some $1\leq s\leq l$, then $\hat{T}_n^{(r_s),r} =O_P(1/n)$, the $s_{th}$
 coordinate of the process on the left side  of   \eqref{revision2-163} degenerates.
 \end{description}
\end{theorem}

A careful inspection of the proof of Theorem \ref{thm6} shows that  \eqref{revision2-163} remains that for any estimator of the change point in the correlation structure that satisfies \eqref{new.116} and \eqref{new.115} (for $\upsilon>1/2$) for any given fixed lag-$k$'s.
Theorem \ref{thm6}  yields an asymptotic level $\alpha$ test for the hypothesis \eqref{hypo3}
of a non-relevant change in the correlation structure  by rejecting $H_0$, whenever
{
\begin{equation} \label{tmax}
\hat
T_{n,\max} :=
\max_{1\leq s\leq l}
 \frac{\hat T^{(r_s),r}_n-\delta^2_{r_s}}{{\color{black}\delta}_{r_s}} ~> { \frac {\bar{v}_{1-\alpha}}{\sqrt{n}}}
\end{equation}
where $\bar v_{1-\alpha}$ denotes
    the $(1-\alpha)$-quantile of the distribution of the random variable
    $$
    \max_{1\leq s\leq l} \{ \mathcal{Z}^{(r_s)}\frac{\Delta_{r_s}}{|\Delta_{r_s}|} \}  ,
    $$
 $\mathcal{Z}^{(r_s)}$  as   defined in \eqref{revision2-163B}. }
   This distribution is a maximum of $l$-variate
    centered normal distributions with a covariance depending on the data generating process in a complicated way, in particular on the long run variance as \eqref{sigma1}.
We construct a bootstrap procedure for generating the critical values with the asymptotically correct nominal level.


Recall the definition
of the  estimator  $\hat t^{(k)}_n$ of the change point in the correlation structure in \eqref{changeestcorr}. Consider the statistics
$$
\hat \Delta^{(k)}_{n,1}={1\over \lf n\hat t^{(k)}_n \rf} \sum_{j=1}^{\lf n\hat t^{(k)}_n \rf}
\frac {\hat e_j \hat e_{j+k}}{\hat \sigma^{2}_n(t_j)} ~,~~
 \hat \Delta^{(k)}_{n,2}={1\over n-\lf n\hat t^{(k)}_n  \rf} \sum_{j=\lf n\hat t^{(k)}_n \rf+1}^{n}
\frac {\hat e_j \hat e_{j+k}}{\hat \sigma^{2}_n(t_j)}
$$
 and take
\begin{equation}  \label{deltadach}
\hat {\Delta}^{(k)}_n=\hat{\Delta}^{(k)}_{n,2}-\hat{\Delta}^{(k)}_{n,1}
\end{equation}
as  an estimator of the difference $\Delta_k=\rho^{(k)}_2-\rho^{(k)}_1$.
We have consistency of $\hat {\Delta}^{(k)}_n$.
(The proof is deferred to
the online supplement.)

\begin{lemma}\label{lemmadelta}
If the conditions of Theorem \ref{thm6} hold, 
  then
$$
\hat{{\Delta}}^{(k)}_n-{\Delta_k}=O_p\big(\tfrac{\log n}{\sqrt{n}}\big)
$$
for $k=r_1,...,r_l$.
\end{lemma}

Let \begin{align}\label{new.new135}
\hat{A}^{(k)}_j={  \frac{\hat{e}_j\hat{e}_{j+k}}{\hat{\sigma}^{2}(t_j)}}-\hat{{\Delta}}^{(k)}_n \mathbf 1(j\geq \lf n\hat t^{(k)}_n \rf),
\end{align}
and let  $\{R_j\}_{j\in \mathbb{Z}}$ be a sequence of  i.i.d.   standard normal distributed random variables independent of $\{\FF_i\}_{i\in\mathbb{Z}}$. We introduce the partial sums
 $\hat{S}_{j,m}^{A,(k)}=\sum_{r=j}^{j+m-1}\hat{A}^{(k)}_r,$ $ \hat{S}_{n}^{A,(k)}=\sum_{r=1}^{n}\hat{A}^{(k)}_r$ and define $\hat{\mathbf S}_{j,m}^{A}=(\hat{S}_{j,m}^{A,(r_1)}...,\hat{S}_{j,m}^{A,(r_l)})$,
  $\hat{\mathbf S}_{n}^{A}=(\hat{S}_{n}^{A,(r_1)}...,\hat{S}_{n}^{A,(r_l)})$,
\begin{align}
\label{107}\hat{\mathbf \Phi}^A_{i,m}=\frac{1}{\sqrt{m(n-m+1)}}\sum_{j=1}^{i}\Big( \hat{\mathbf S}_{j,m}^A-\frac{m}{n}\hat{\mathbf S}_n^A \Big)R_j.
\end{align}
Let  $\hat{\Phi}^{A,(s)}_{i,m}$ be the  $s_{th}$ component of $\hat{\mathbf \Phi}^A_{i,m}$.
Then the following result is proved in Section \ref{sec63} of the  online supplement.
\begin{theorem}
\label{thm7}
Suppose the conditions of Theorem \ref{thm6} hold and that $m\rightarrow \infty$, $m\log n/\sqrt n\rightarrow 0$,
 { $\sqrt{m}\big(c_n^2+(\frac{1}{\sqrt{nc_n}}+b_n^2+\frac{1}{\sqrt{nb_n}})c_n^{-1/4}\big)\log n\rightarrow 0$}. If for $1\leq s\leq l$,
\begin{align*}M_n^{r,(r_s)}=
\frac{1}{n} \frac{6  }{(\hat t^{(r_s)}_n)^2(1-\hat t^{(r_s)}_n )^2}
 \sum_{m+1\leq i\leq n-m+1}\Big(\hat{\Phi}^{A,(s)}_{i,m}-\frac{i}{n-m+1}\hat{\Phi}^{A,(s)}_{n-m+1,m}\Big)\Big(\frac{i\hat t^{(r_s)}_n}{n}-\frac{i}{n}\wedge\hat t^{(r_s)}_n\Big)
 \end{align*}
 then
 (conditional on $\FF_n$ in probability)
\begin{equation}\label{bootcorrel}
(M_n^{r,(r_1)}...,M_n^{r,(r_l)})^T\cod  { \tilde {\mathcal{Z}} :=
 \{\mathcal{Z}^{(r_s)} \}_{s=1}^l~}
\end{equation}
 where the random variables $\{\mathcal{Z}^{(r_s)} \}_{s=1}^l~$ are defined in Theorem \ref{thm6}.
\end{theorem}

The bootstrap  test  for the hypothesis \eqref{relvar}  of a  non-relevant change   in the correlation structure results as follows. 
\begin{algorithm}\label{algorithmcorrel} ~~\\ {\rm
[1] Calculate the statistics  $ \hat{T}^{(r_u),r}_n $ defined in (\ref{76}) for $u=1, \ldots , l$. For given $\delta=(\delta_{r_1}...,\delta_{r_l})^T$, calculate $\hat  T_{n,\max}$ by \eqref{tmax}.

\noindent
[2] Generate $B$  conditionally $i.i.d$ copies $\{\hat{\mathbf \Phi}^{A}_{i,m,r}\}_{i=1}^{n-m+1}$ ($r=1,2,...,B$)
of the sequence $\{\hat{\mathbf \Phi}^A_{i,m}\}_{i=1}^{n-m+1}$ defined in \eqref{107}.
 Calculate $M_r^A:=\max_{1\leq u\leq l}(M_{n,r}^{A,(r_u)})$ where, for $1\leq u\leq l$,
    $$
    M_{n,r}^{A,(r_u)}=\frac{1}{n} \frac{ 6 {sign}(\hat\Delta^{(r_u)}_{n}) }{(\hat t^{(r_u)}_n)^2(1-\hat t^{(r_u)}_n)^2}  \sum_{i=m+1}^{n-m+1}\Big(\hat{\Phi}_{i,m,r}^{A,(u)}-\frac{i}{n-m+1}\hat{\Phi}^{A,(u)}_{n-m+1,m,r}\Big)\Big(\frac{i\hat t^{(r_u)}_n}{n}-\frac{i}{n}\wedge\hat t^{(r_u)}_n \Big).
    $$
[3] If $M_{(1)}^A\leq M_{(2)}^A\leq ... \leq M_{(B)}^A$ denote the order statistics of $M_1^A, \ldots , M_B^A$, reject the null hypothesis \eqref{relvar} of a non-relevant change in the correlations  at level $\alpha$ if
{
     \begin{equation} \label{bootrel}
\hat  T_{n,\max}  >  \frac{M_{(\lf B(1-\alpha)\rf)}^A }{\sqrt{n}}.
    \end{equation}
 The $p$-value of this test is given by $1-\frac{B^*}{B}$,  where  $B^*=\max\{r: \frac{M^A_{(r)}}{\sqrt{n}}\leq \hat T_{n,\max}\}$.}
 }
\end{algorithm}
If only one lag is considered, then the term ${sign}(\hat\Delta^{(r_u)}_{n})$  in the definition of $M^{A,(r_u)}_{n,r}$ can be dropped by the symmetry of a centered Gaussian process.

\begin{remark}\label{discusrelevant}   { \rm
We investigate the power of the test \eqref{bootrel}. Let $\bar v_{r_s,1-\alpha}$ be the $(1-\alpha)$-quantile of
the distribution of the random variable $\max\{\mathcal{Z}^{(r_s)}sign(\Delta_{r_s})\}_{s=1}^l$.
If $\Delta_{r_1}^2 > \delta^2_{r_1}$, then we obtain from Theorem \ref{thm6} an approximation for the
power of the test \eqref{tmax} as
 \begin{align} \nonumber
\beta_n (\delta,\Delta) & :=  \p  \Big( \hat  T_{n,\max} >  \frac{\bar v_{1-\alpha}}{\sqrt{n}}\Big)
\geq \p\Big(\hat T^{(r_1),r}_n>\delta_{r_1}^2+\frac{\bar v_{1-\alpha}\delta_{r_1}}{\sqrt{n}}\Big)
\notag\\
& \nonumber
=\p\Big(\sqrt{n}\frac{\hat T^{(r_1),r}_n-\Delta_{r_1}^2}{ {   |\Delta_{r_1} |}   }>\sqrt{n}\frac{\delta_{r_1}^2-\Delta_{r_1}^2}{  {   |\Delta_{r_1} |}  }+\frac{\bar v_{1-\alpha}\delta_{r_1}}{  {   |\Delta_{r_1} |}  }\Big) \\
&\approx  1 - \Psi_{r_1} \Bigl( \sqrt{n}\frac{\delta_{r_1}^2-\Delta_{r_1}^2}{ {   |\Delta_{r_1} |}  } +  \frac{\bar v_{1-\alpha}\delta_{r_1}}{ {   |\Delta_{r_1} |}  } \Bigr) ,
\label{approx}
\end{align}
where $\Psi_{r_1}$ is the distribution function of the random variable $ {\cal Z}^{(r_1)} $  (in fact a centered normal distribution).
Therefore,  under the  alternative of a relevant change for some lag $r_1$, $\Delta_{r_1}^2 > \delta_{r_1}^2 $, we have  $ \beta_n(\delta,\Delta) \to 1 $ as $n\to \infty$, which
provides the consistency of the  test \eqref{bootrel}. Under the null hypothesis $0< \Delta_{r_s}^2 \leq   \delta_{r_s}^2 $ for $1\leq s\leq l$,
we have
{
\begin{align}\label{new4.15}
1-\beta_n (\delta,\Delta)&=  \p  \Big( \hat  T_{n,\max} \leq   \frac{\bar v_{1-\alpha}}{\sqrt{n}}\Big)
\\&= \p\Big(\max_{1\leq s\leq l}\Big \{\frac{\Delta_{r_s}}{\delta_{r_s}}\mathcal {Z}^{(r_s)}+\sqrt{n}\frac{\Delta_{r_s}^2-\delta^2_{r_s}}{\delta_{r_s}}\Big \}\leq \bar v_{1-\alpha}
\Big) \big( 1+ o(1) \big)
\end{align}
Consequently, if $0< \Delta_{r_s}^2 \leq   \delta_{r_s}^2 $  ($1\leq s\leq l$) and
$$ if
\l^* := \# \big\{ s \in \{1,\ldots ,l\} ~|~|\Delta_{r_s}| = \delta_{r_s} \big\}
$$
denotes the number of coordinates where the ``true''  difference between the lag-$r_s$ correlations is  at the boundary of the null  hypothesis,
{\color{black} we have
\begin{eqnarray} \label{level}
\lim_{n\to \infty } \beta_n (\delta,\Delta)   \left\{ \begin{array}{ll}
=0 &  \mbox{ if } l^*=0 \\
=\alpha & \mbox{ if }  l^*=l \\
<\alpha& \mbox{ if }   1 \leq l^*\leq l-1.
\end{array}
\right.
\end{eqnarray}}
\noindent If there exist some lags, without loss of generality $r_1,\ldots ,r_k$,
with $\Delta_{r_i}=0$ ($1\leq i\leq k$) and  $k<l$, then it follows  that $\hat{T}^{(r_i),r}_n =O_P(1/n)$ for all $1\leq i\leq k$, and
it is easy to see that a result similar to \eqref{level} holds. Moreover,
if  $\Delta_{r_s}=0$ for all $s= 1, \ldots , l$, then $\hat{T}^{(r_s),r}_n =O_P(1/n)$
for $s= 1, \ldots , l $ and  $\lim_{n\to \infty } \beta_n (\delta,\Delta)  = 0$ (since  $\alpha\leq 0.5$ and $\bar v_{1-\alpha}>0$).
Summarizing these calculations shows that the test \eqref{bootrel} has, in fact, asymptotic level $\alpha$.  \\
We can also use \eqref{approx}  to investigate the power as a function of the parameter $\delta$
in the hypothesis \eqref{relvar}: for
sufficiently large  $n$  the power
$\beta_n (\delta,\Delta) $ is approximately $1$  if $\delta\rightarrow 0$ and $\sqrt n \delta\rightarrow \infty$, and approximately $0$
if $\delta\rightarrow \infty$. Moreover,
it is  easy to see that all   statements mentioned in this remark  hold also for the  bootstrap test defined by \eqref{bootrel}.}
}
\end{remark}

\begin{remark}  {\rm  \color{black} \label{biascor}
In applications of the  test \eqref{bootrel} for a non-relevant change  in the correlation, the thresholds $\Delta_{r_s} $ are usually very small, and this can lead to
a less accurate approximation of the nominal level. Consider, for example, the univariate test  for a relevant change in the lag-$1$ correlation. We obtain from the proof of
 in Theorem \ref{thm6}
for the {\color{black} estimating object} of statistic defined in \eqref{76} the stochastic expansion {\color{black}(omitting the subscript)}
\begin{align}
\sqrt n(T_n^2-\Delta^2)&=\frac{6\Delta}{t^2(1-t)^2}\int (U(s)-sU(1))(st-s\wedge t)ds\notag
\\&+\frac{3}{\sqrt n t^2(1-t)^2}\int (U(s)-sU(1))^2ds+o_p(n^{-1/2}),\label{finit-explain}
\end{align}
where $t$ is the jump time in lag-$1$ correlation and the process  $\{U(t)\}_{t\in [0,1]}$ is defined in Theorem \ref{theorem1}.
The second term vanishes asymptotically. However, when $\Delta$ is small and the sample size is not too  large, the first and second term on the
right hand side   of \eqref{finit-explain} could be comparable in size. The bootstrap methodology
proposed in this paper provides us with  a convenient way to solve this problem.  We  propose  to replace $\hat  T_{n,\max}$ in \eqref{tmax} by $\max_{1\leq u\leq l}\{\hat{T}^{(r_u),r}_n-\delta^2_{r_u}\}$, and to replace the statistic $    M_{n,r}^{A,(r_u)}$
in step [2] of Algorithm \ref{algorithmcorrel} by the statistic
    $$
    M_{n,r}^{A,(r_u)}=\frac{1}{n} \frac{ 6 {sign}(\hat\Delta^{(r_u)}_{n})\delta_{r_u} }{(\hat t^{(r_u)}_n)^2(1-\hat t^{(r_u)}_n)^2}  \sum_{i=m+1}^{n-m+1}\Big(\hat{\Phi}_{i,m,r}^{A,(u)}-\frac{i}{n-m+1}\hat{\Phi}^{A,(u)}_{n-m+1,m,r}\Big)\Big(\frac{i\hat t^{(r_u)}_n}{n}-\frac{i}{n}\wedge\hat t^{(r_u)}_n \Big)+N^A_{n,r_u}
    $$
    where
    $$
    N_{n,r}^{A,(r_u)}=\frac{1}{n^{3/2}} \frac{ 3}{(\hat t^{(r_u)}_n)^2(1-\hat t^{(r_u)}_n)^2}  \sum_{i=m+1}^{n-m+1}\Big(\hat{\Phi}_{i,m,r}^{A,(u)}-\frac{i}{n-m+1}\hat{\Phi}^{A,(u)}_{n-m+1,m,r}\Big)^2.
    $$
}
\end{remark}
}
\begin{remark} {\rm \label{newremark4.3}
Straightforward calculation shows that the computational time complexity of Algorithms \ref{algvar} and \ref{algorithmcorrel} is $O(Bn+\alpha(n))$, where $n$ is the length of time series, $\alpha(n)$ is the time cost of obtaining $\{\hat e_i\}_{1\leq i\leq n}$ and $\{\hat \sigma^2(t_i)\}_{1\leq i\leq n}$ which depends on the particular optimization method that users choose, and $B$ is the number of bootstrap replications that  is mainly determined by the nominal level. As a rule of thumb, for a nominal level of 5\%, our experience shows that $B=2000$ is sufficient, though we use $B=4000$ and $8000$ in our simulations and data analysis, respectively.}
\end{remark}

\begin{remark} {\rm
\label{sec:newremark4.4}

For the change point test defined in Algorithm \ref{algvar} the alternative hypothesis allows for multiple change points and one could use a similar approach as in Section 5 of \cite{qu2008}. For the test
of  relevant change points defined in Algorithm \ref{algorithmcorrel} we propose to proceed in two steps: we   use Algorithm \ref{algvar} and
the binary segmentation technique to deal with  multiple change points (see \cite{vostrikova1981}); if this procedure identifies the potential relevant change points  $0=t_0<t_1<, \ldots  <t_s<t_{s+1}=1$,   we perform a test for a  relevant change point  in every two consecutive intervals $(t_l,t_{l+2}]$ for $0\leq l\leq s-1$.

 }
\end{remark}

\begin{remark}\label{new.remark.4.5}{\rm
	The  behaviour of the test statistics may not be close to the limiting distribution when the sequence is short, especially under piecewise local stationarity.  As a result,  the finite performance of  those tests only  based on the limiting distribution may not be satisfactory under non-stationarity. Thanks to the bootstrap procedure, our proposed method works reasonably well  and is not very sensitive to the length of the sequence. This is also justified by the simulation results for sample sizes 300, 500, 800 in Section \ref{differen-sample-simulation} of the supplementary material. As a rule of thumb, we recommend our method when the length of sequence is larger than 300.
}
\end{remark}

\section{Finite sample properties}   \label{sec5}
\def\theequation{5.\arabic{equation}}
\setcounter{equation}{0}

In this section we investigate the finite sample properties of the proposed tests by means of a simulation study.
In all examples considered   we used the quadratic mean function
 $\mu(t)=8(-(t-0.5)^2+0.25)$   and a sequence of independent
 identically random variables   $\{\varepsilon_j\}_{j\in \mathbb{Z}}$ in the definition
 of the errors  $e_i=G_j(t_i,\FF_i)$ in model \eqref{mod1} where $\FF_ i = \sigma (\ldots, \varepsilon_0, \ldots, \varepsilon_i)$, if not mentioned otherwise.
 The dependence structures differ by choice of the nonlinear filters $G_j$.
 The sample size was  $n=500$ and all results were based on $4000$ simulation runs. In each run,
 the critical values were generated by $B=4000$ bootstrap replications. We use  the Epanechnikov kernel;
 { We analyzed the impact of  different kernel functions on the performance of the tests and saw no substantial differences. Some of these investigations are summarized
  in Section \ref{different-kernel} of the supplementary material.}

\begin{figure}[t]
\centering
  \includegraphics[width=13cm,height=8cm]{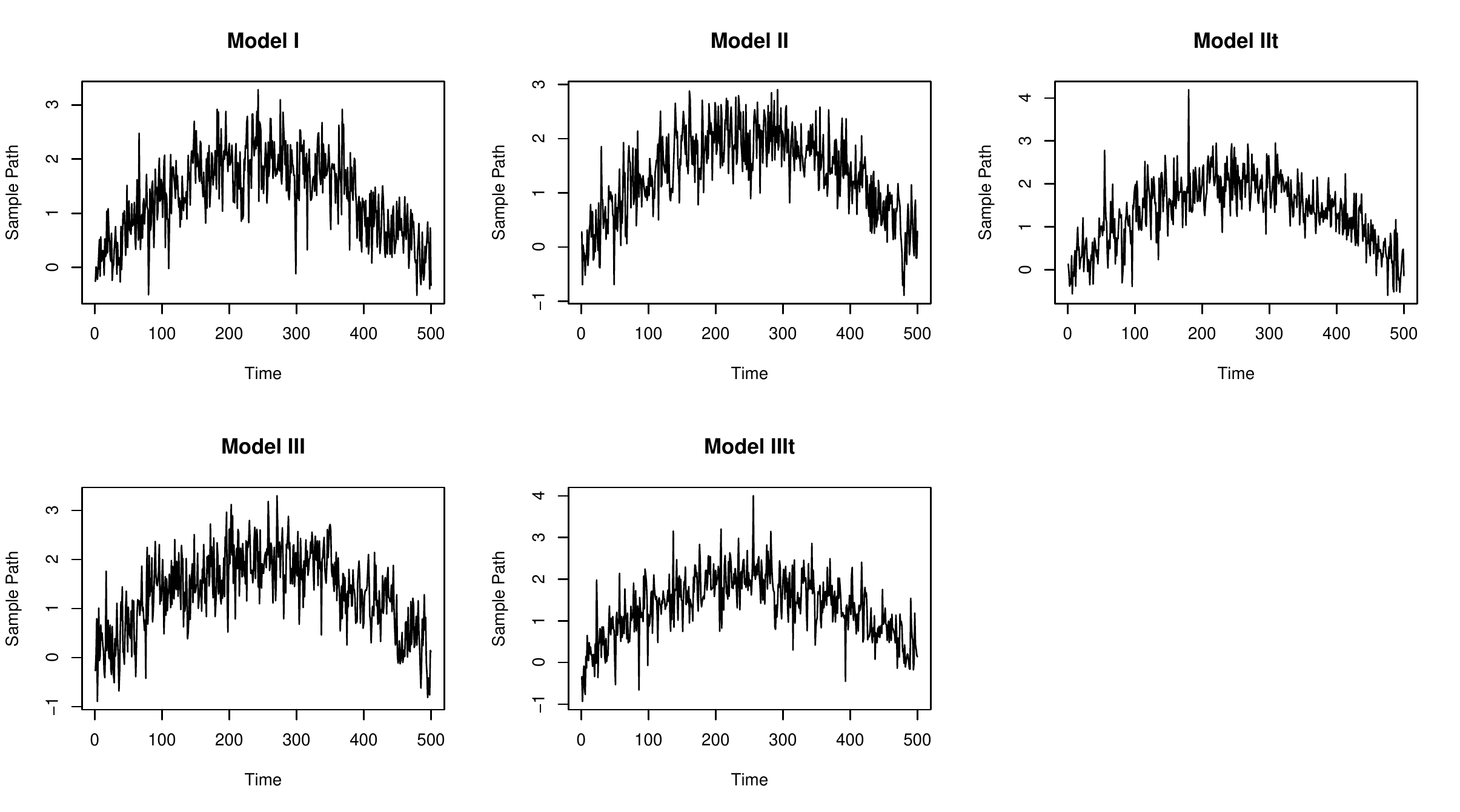}
\caption{\it Typical sample paths of the processes corresponding to model  $(I)$ - $(IV)$.}
  \label{fig:mini:subfig:fige01}
\end{figure}

\subsection{Change point tests for correlations}\label{sec:5.1}
{
 We   investigate  properties of the tests for  changes in the lag-$1$ and lag-$2$ correlations. For this purpose we consider these models.
 \begin{description}
\item (I)    $G(t,\FF_i)=H(t,\FF_i)\sqrt{1- (t-0.5)^2}/2$, where $H(t,\FF_i)=0.2H(t,\FF_{i-1})+\varepsilon_i$.

\item (II,IIt)  $G(t,\FF_i)=H(t,\FF_i)\sqrt{c(t)}/2$ for $t\leq 0.5$, and $G(t,\FF_i)=H(t,\FF_i)\sqrt{d(t)}/2$ for $t>0.5$, where
$c(t)=1- (t-0.5)^2$, $d(t)=1-\frac{1}{2}\sin t$, and  $H(t,\FF_i)=0.2H(t,\FF_{i-1})+\varepsilon_i$.
\item (III,IIIt)  $G(t,\FF_i)=H(t,\FF_i)\sqrt{1- (t-0.5)^2}/2$, where $H(t,\FF_i)=0.1H(t,\FF_{i-1})+\varepsilon_i$
for $t\leq 0.5$, and $H(t,\FF_i)=0.4H(t,\FF_{i-1})+\varepsilon_i$ for $t>0.5$.
\item (IV) $G(t,\FF_i)=H(t,\FF_i)\sqrt{1- (t-0.5)^2}/2$, where $H(t,\FF_i)=0.5H(t,\FF_{i-1})+0.1H(t,\FF_{i-2})+\varepsilon_i$
for $t\leq 0.5$, and $H(t,\FF_i)=0.3H(t,\FF_{i-1})+0.2H(t,\FF_{i-2})+\varepsilon_i$ for $t>0.5$.
\end{description}
For models (I) (II) (III) (IV) the innovations were $\varepsilon_i\sim_{i.i.d}\ N(0,1)$, and for  model (IIt) and (IIIt)
,  $\varepsilon_i\sim_{i.i.d}\ {t(5)}/{\sqrt{5/3}}$. Model (I) was for locally stationary processes. The variance of the process was time-varying, but the correlation remained constant. Model (II,IIt),
  (III,IIIt) and model (IV) were piecewise locally stationary processes, where the variances  had an abrupt change. Before and after the jump, the variance varied smoothly.
 The correlations of model  (I) and (II,IIt) were constant, while the correlations of model (III,IIIt,IV) had a break at $t=0.5$ and  were   used to
 illustrate the approximation of the nominal level of the test for the
 hypothesis of a  non-relevant change point, as discussed at the end of this section. Model (IV) is a tvAR($2$) model with a change in the lag-$1$ and lag-$2$ scaled AR coefficients.
  Typical trajectories corresponding to these processes are depicted in  Figure \ref{fig:mini:subfig:fige01}

Change point analysis on the basis of the tests proposed in Section \ref{sec3} and \ref{sec4} requires the choices of two bandwidths
 in the local
linear estimates of the mean and variance.
We  used a generalized cross validation method (GCV) introduced by \cite{zhouwu2010}
to select the bandwidth for estimating the mean function. Then we applied this cross validation procedure
 again to select the bandwidth for estimating the variance function. The parameters $L$ and $\zeta$ in the estimator
 \eqref{tstar} were chosen as $L= \lfloor 3n^{1/3}\rfloor $ and $\zeta = 0.2$, respectively. For the choice of window size $m$ in Section \ref{sec3} and \ref{sec4} we used the minimal volatility method (MV) in \cite{zhou2013}.

For the nominal level we display in Table \ref{tab:addlabel2}
the rejection probabilities of the test for the hypothesis \eqref{mulcorr0}
 of a ``classical''  change point, where  various bandwidths $b_n$ from  the interval $ [0.075,0.225]$ were considered.
 At each fixed $b_n$, the bandwidth $c_n$ for estimating the  variance was calculated by cross validation.
 The last row of the table  shows the simulated rejection probabilities for the case that both  bandwidths $b_n$ and $c_n$
were   calculated by cross validation.  In the $1_{st}$-$3_{rd}$ column  we display results of the test \eqref{testvar} for models I, II and IIt, where we used
lag-$1$ correlation.  The $4_{th}$  column  (denoted by II$^*$)  corresponds to  model II,  where  lag-$1$ and lag-$2$ correlations were used
  simultaneously in the test \eqref{testvar}.  We observed a reasonable approximation of the nominal level, only slightly affected by the choice of the bandwidth
  $b_n$. Moreover, generalized  cross validation yielded a good approximation of the nominal level in all cases under consideration.
  
  In Table \ref{tab:addlabel2rel}  we show corresponding results for the test  \eqref{bootrel}  of  a non-relevant change point, where
  in all cases the simulated type I error was calculated for a boundary point of the null hypothesis. Thus $l^* =l$ in \eqref{level}
  and, by the discussion in Remark \ref{discusrelevant}  the  nominal level  of the test should be close to $\alpha$  at this point.
In the $1_{st}$ and $2_{nd}$ columns we show  the simulated type I error
 of the test  \eqref{bootrel} for a relevant change in lag-$1$ correlation with  $\delta=\Delta=0.3$ for Models III and IIIt, respectively. In the $3_{rd}$ column of Table \ref{tab:addlabel2rel}
 we display  the simulated level of the test for the hypotheses  \eqref{hypo3} for a relevant change in lag-$1$ and lag-$2$ correlations for model III,
 where  $\delta_1=\Delta_1=0.3$ and $\delta_{2}=\Delta_2=0.15$, respectively. Finally, the $4_{th}$ column shows corresponding results for the locally stationary AR(2) model
 (IV) where again  lag-$1$ and lag-$2$ correlations were considered (here $(\delta_1,\delta_2)=(\Delta_1,\Delta_2)=(0.18,0.065)$).
Once again, all displayed results correspond to the boundary, and at interior points of the null hypothesis the type I error of the test \eqref{bootrel} is usually smaller
 (see the discussion in Remark  \ref{discusrelevant}).
 }

\begin{table}[htbp]
  \centering
  \caption{\it {Simulated Type I error of the test for the classical hypothesis \eqref{mulcorr0}
of   a change in the correlation for various bandwidths and the  bandwidth  calculated by generalized cross validation (last line).
 Columns $1-3$: test  \eqref{testvar} based on the lag-$1$ correlation for Models I, II and IIt. Column $4$:  test   \eqref{testvar} based on lag-$1$ and  lag-$2$  correlations for Model II. }}
    \begin{tabular}{r|rrrrrrrr}
    \toprule
    model & \multicolumn{2}{c}{I} & \multicolumn{2}{c}{II} & \multicolumn{2}{c}{IIt} & \multicolumn{2}{c}{II*} \\
    \midrule
    $b_n/\alpha$     & 5\%   & 10\%  & 5\%   & 10\%  & 5\%   & 10\%  & 5\%   & 10\% \\
    \hline
    0.075 & 5.625  & 11.6   & 4.375  & 9.6    & 4.825 & 10.025 & 4.725  & 10.35 \\
    0.1   & 5.2    & 10.8   & 4.3    & 9.775  & 4.9   & 10.925 & 5.325  & 10.675 \\
    0.125 & 4.025  & 9.35   & 4.05   & 9.275  & 4.075 & 8.875  & 4.425  & 8.95 \\
    0.15  & 4.575  & 10.075 & 3.75   & 8.4   & 4.35  & 9.75   & 3.95   & 9.45 \\
    0.175 & 4.1    & 8.675  & 3.85   & 8.75   & 3.65  & 8.4    & 4.175  & 9.05 \\
    0.2   & 3.725  & 8.6    & 3.575  & 8.15   & 3.525 & 8.225  & 3.825  & 8.35 \\
    0.225 & 3.925  & 8.675  & 3.2    & 8.025  & 4.025 & 8.65   & 3.95   & 8.625 \\
        \hline
    GCV    & 4.275 & 9.625   & 4.575 & 9.425 & 4.25 & 9.475  & 4.75  & 9.8 \\
    \bottomrule
    \end{tabular}%
  \label{tab:addlabel2}%
\end{table}%

\begin{figure}[htbp]
\centering
  \includegraphics[width=6.5cm,height=6.5cm]{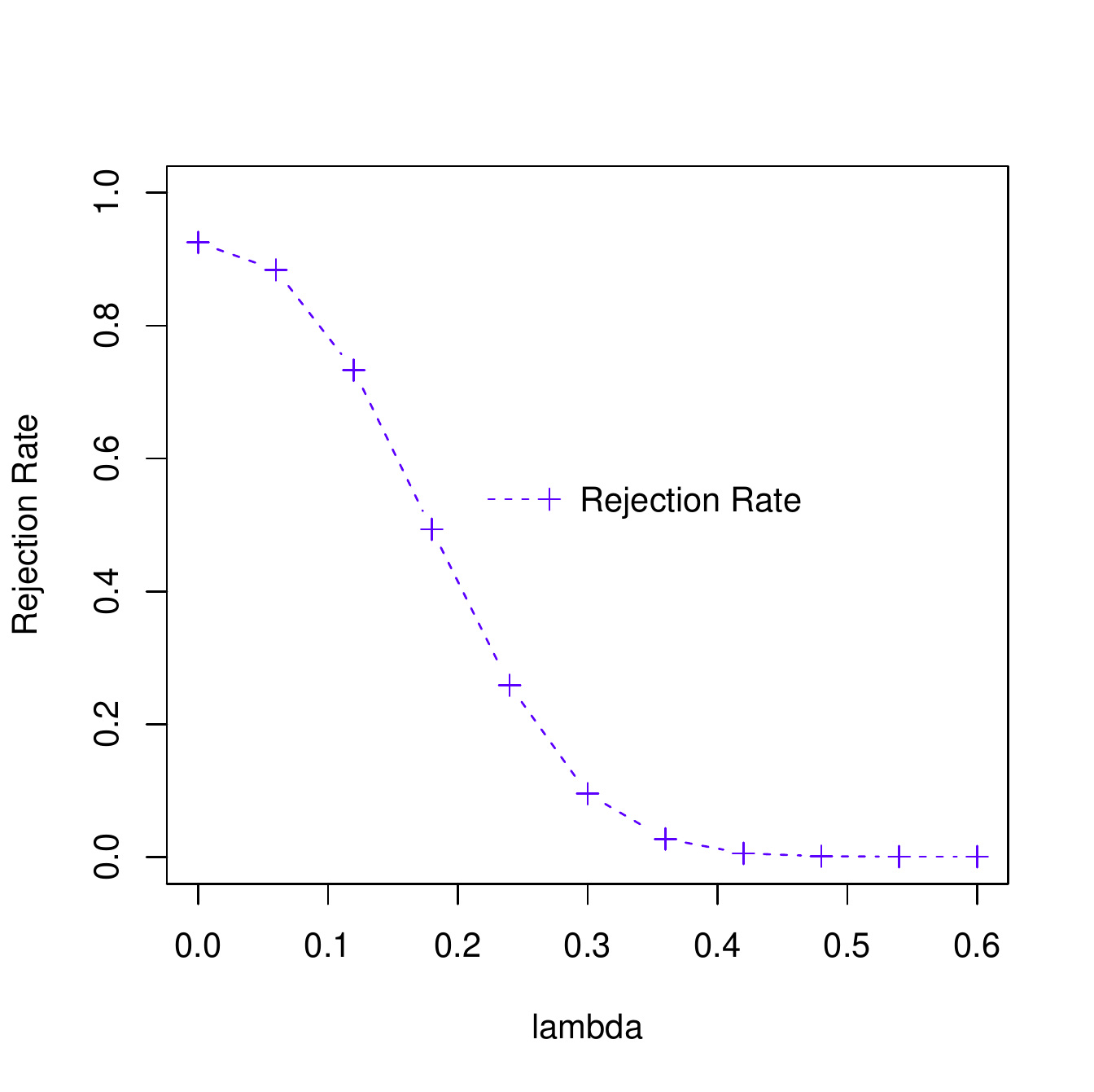}
\caption{\it Simulated rejection probabilities of the test for lag-$1$ correlation 
as a function of the threshold $\delta \in[0,2\Delta]$ in the hypothesis \eqref{relvar} for Model III.}
  \label{fig:mini:subfig:fige03}
\end{figure}

\begin{table}[htbp]
  \centering
  \caption{\it {Simulated Type I error of the test for the hypothesis \eqref{hypo3}
of   a relevant change in the correlation for various bandwidths and the  bandwidth  calculated by generalized cross validation (last line).
 Columns $1$ and $2$: tests based on the lag-$1$ correlation for Models III and IIIt. Column $3$ and $4$:  test based on lag-$1$ and  lag-$2$  correlations for Models III and IV. }}
     \begin{tabular}{r|rrrrrrrr}
    \toprule
    model & \multicolumn{2}{c}{III} & \multicolumn{2}{c}{IIIt} & \multicolumn{2}{c}{III*} & \multicolumn{2}{c}{IV} \\
    \midrule
   $ b_n$/$\alpha $ & 5\%   & 10\%  & 5\%   & 10\%  & 5\%   & 10\%  & 5\%   & 10\% \\
       \hline
    0.075 & 5.275  & 9.575   & 6.65  & 10.775 & 4.925   & 9.9      & 5.6    & 10.85 \\
    0.1   & 6.2    & 10.825  & 6.325 & 10.575 & 5.45    & 10.475   & 5  & 9.825 \\
    0.125 & 6.05   & 11.3    & 6.425 & 11.125 & 5       & 10.025   & 4.875 & 9.55 \\
    0.15  & 5.8    & 10.25   & 6.575 & 11.075 & 5.25    & 10.35    & 4.025  & 8.325 \\
    0.175 & 5.775  & 10.1    & 6.075 & 10.8   & 4.85     & 10.275  & 4.175  & 8.5 \\
    0.2   & 5.775  & 9.425   & 5.575 & 9.9    & 4.775       & 10.1 & 4      & 8.925 \\
    0.225 & 5.3    & 10.15   & 5.45  & 9.95   & 4.525   & 9.425    & 3.75  & 8.2 \\
        \hline
    GCV    & 5.45 & 9.9 & 5.875 & 10.55 & 5.45 & 10.675 & 4.875  & 9.6 \\
    \bottomrule
    \end{tabular}%
  \label{tab:addlabel2rel}%
\end{table}%
Figure  \ref{fig:mini:subfig:fige03} shows the simulated rejection probabilities
of the tests for the hypothesis  \eqref{hypo3}  of a non-relevant change in the  lag-$1$ correlation
for model III as a function of the parameter   $\delta \in[0,2\Delta] $.
 The significance level was chosen as  $0.1$. As expected the probability of rejection
 decreases with $\delta$ (see also the discussion in Remark \ref{discusrelevant}).
{ More simulation results for different sample sizes can be found in  Section \ref{differen-sample-simulation} of the supplementary material.}

{
Importantly, the symmetry of the innovations do not affect the  asymptotic properties
of the tests, since the rates of Gaussian approximations of  partial sums from skewed random variables are of
 the same order as in the symmetric case. To investigate
if there exist
differences in the  finite sample properties
 we took model II and III, with the  $i.i.d.$ Gaussian innovations 
replaced by $i.i.d$ $(\chi^2(5)-5)/\sqrt{10}$ random variables.  In Table \ref{tab:asymmetric}
we display the
simulated type I error of  the  test  \eqref{testvar}
for a change point  in the  lag-$1$ correlation in  model $II$ and of the test
\eqref{bootrel}  for a relevant change point  in the lag-$1$ correlation in model $III$. The corresponding results for a symmetric error can be found in Tables
\ref{tab:addlabel2} and \ref{tab:addlabel2rel}; we only  observe minor differences in the approximation of the nominal
level between the symmetric and non-symmetric case.
\begin{table}[htbp]
	\centering
	\caption{\it Simulated type I error
	of  the  test  \eqref{testvar} for a change point  in the  lag-$1$ correlation in
	 model $II$ and of the test
\eqref{bootrel}  for a relevant change point test in lag-$1$ correlation (at the  boundary point of the null) in model $III$ with $(\chi^2(5)-5)/\sqrt{10}$ innovations. The last column
represents the simulated Type I error if the bandwidth is  $b_n$ selected by GCV.
	}
	\begin{tabular}{crrrrrrrrr}
		\toprule
		\multirow{3}[0]{*}{~~~~~~ II } model & $b_n$  & 0.075 & 0.1   & 0.125 & 0.15  & 0.175 & 0.2   & 0.225 & GCV \\
		\midrule
		& 5\%   & 4.6   & 4.55  & 3.6   & 2.6   & 2.95  & 3.25  & 2.9   & 4.2 \\
		& 10\%  & 10.35 & 8.95  & 8.7   & 7.15  & 7.65  & 7.35  & 6.8   & 9.3 \\
		\multirow{2}[0]{*}{ III } & 5\%   & 4.85  & 5.1   & 4.95  & 6.9   & 5.25  & 6     & 5.35  & 5.4 \\
		& 10\%  & 8.9   & 8.85  & 9.7   & 11.7  & 9.9   & 10.35 & 9.85  & 9.95 \\
		\bottomrule
	\end{tabular}%
	\label{tab:asymmetric}%
\end{table}%
}

\begin{remark}{\rm \label{Remark5.1-new}
{
	Throughout, the bandwidth  $b_n$ is assumed to be the same over the whole sequence.
	As pointed out by one referee, it might be of interest to investigate a  time-dependent bandwidth  $b_n$  with
	respect to its  potential to deal with local stationarity. Using similar arguments as in  \cite{zhouwu2009}  we can obtain the optimal time varying bandwidth
	as 	
	\begin{equation} \label{varband}
	b_n(t)=\left|\frac{\hat \kappa(t)}{\hat \sigma(t)}\right|b_n,
	\end{equation}
	where $b_n$ is the time invariant bandwidth obtained by the GCV method, $\hat \kappa^2$ and $\hat \sigma^2$ are estimates of the long run variance and the variance of the random variables $e_i$, respectively. It is hard to accurately estimate $\kappa^2$ in  a PLS
	model due to the unknown break points. In the case of  local stationarity,  an estimate of $\kappa^2$ was proposed by
	\cite{zhouwu2010} and we used this method to investigate the differences between a local and global bandwidth
in the locally stationary  model I.  The simulated levels of the corresponding bootstrap tests are shown in Table \ref{Table:Varying} and we observe
	that the performance of the procedure with a time dependent  bandwidth  is quite similar to the one using a constant bandwidth.
\begin{table}[htbp]
	\centering
	\caption{\it Simulated type I error
	of  the  test  \eqref{testvar} for a change point  in the  lag-$1$ correlation in
model I using the time varying bandwidth \eqref{varband}. The last column
represents the simulated Type I error if the bandwidth is  $b_n$ selected by GCV.}
	\begin{tabular}{crrrrrrrrr}
		\toprule
		\multirow{3}[0]{*}{Model I} & $b_n$  & 0.075 & 0.1   & 0.125 & 0.15  & 0.175 & 0.2   & 0.225 & GCV \\
		\midrule
		& 5\%   & 5.05  & 4.9   & 4.3   & 3.95  & 4.2   & 3.15  & 3.55  & 4.15 \\
		& 10\%  & 10.45 & 9.75  & 9.5   & 9.85  & 8.9   & 7.55  & 8     & 10.7 \\
		\bottomrule
	\end{tabular}%
	\label{Table:Varying}%
\end{table}%
}	}
\end{remark}

{
\subsection{Some robustness considerations}\label{Non-PLS-Simulation}
 As  was pointed out by a referee it might be of interest to investigate the approximation
  of the  nominal level if the assumption of PLS is violated. For this purpose we considered modifications of the models $II$ and $III$ introduced in the  previous section. Let
  $\{\eta_i, i\in \mathbb Z\}$  denote  $i.i.d.$  standard normal distributed
  and $\{\varepsilon_i,i\in \mathbb Z\}$ denote  $i.i.d$  $t$-distributed random variables with
  $5$ degrees  of freedom, normalized such that they have variance $1$. We consider the processes
\begin{description}
	\item ($II_0$) $G_i=H_i\sqrt{c(i/n)}/2$ for $i/n\leq 0.5$, and $G_i=H_i\sqrt{d(i/n)}/2$ for $i/n>0.5$, where
	$c(t)=1- (t-0.5)^2$, $d(t)=1-\frac{1}{2}\sin t$ and  $H_i=0.2H_{i-1}+\varepsilon_i$ for $i/n\leq 0.5$, and $H_i=0.2H_{i-1}+\eta_i$ for $i/n>0.5$.
	\item ($III_0$) $G_i=H_i\sqrt{1- (i/n-0.5)^2}/2$, where $H_i=0.1H_{i-1}+\varepsilon_i$
	for $i/n\leq 0.5$, and $H_i=0.4H_{i-1}+\eta_i$ for $i/n>0.5$.
\end{description}
These models are not PLS in the sense of  Definition \ref{def1}.
In Table \ref{tab:non_PLS} we show the simulated type I error of  the  test  \eqref{testvar}
for a change point  in the  lag-$1$ correlation in  model $II_0$ and of the test
\eqref{bootrel}  for a relevant change point test in the  lag-$1$ correlation in model $III_0$.
We observe reasonable approximations of the nominal level in all  cases under consideration.
\begin{table}[htbp]
	\centering
	\caption{\it  Simulated type I error of  the  test  \eqref{testvar}
for a change point  in the  lag-$1$ correlation in  model $II_0$ and of the test
\eqref{bootrel}  for a relevant change point test in the lag-$1$ correlation  (at the
 boundary point of the null) in  model $III_0$.
 The last column
represents the simulated Type I error if the bandwidth is  $b_n$ selected by GCV.}
	\begin{tabular}{crrrrrrrrr}
		\toprule
		\multirow{3}[0]{*}{$II_0$} & $b_n$  & 0.075 & 0.1   & 0.125 & 0.15  & 0.175 & 0.2   & 0.225 & GCV \\
		\midrule
		& 5\%   & 4.45  & 3.9   & 3.8   & 4.15  & 2.9   & 3     & 2.85  & 3.6 \\
		& 10\%  & 9.8   & 9.2   & 9.2   & 8.4   & 7     & 7.55  & 6.7   & 8.55 \\
		\multirow{2}[0]{*}{$III_0$}
		& 5\%   & 6.25  & 5.75  & 5.9   & 5.2   & 6.2   & 6.05  & 4.7   & 4.6 \\
		& 10\%  & 10.15 & 10    & 10.1  & 9.7   & 11.4  & 10.8  & 8.8   & 8.35 \\
		\bottomrule
	\end{tabular}%
	\label{tab:non_PLS}%
\end{table}%
In fact, it follows from \cite{Zhou2012} that model $II_0$ and $III_0$ can be approximated by  the  two PLS models
\begin{description}
	\item ($II_0^*$). $G(t,\FF_i)=H(t,\FF_i)\sqrt{c(t)}/2$ for $t\leq 0.5$, and $G(t,\FF_i)=H(t,\FF_i)\sqrt{d(t)}/2$ for $t>0.5$, where
	$c(t)=1- (t-0.5)^2$, $d(t)=1-\frac{1}{2}\sin t$ and  $H(t,\FF_i)=0.2H(t,\FF_{i-1})+\varepsilon_i$ for $t\leq 0.5$, and  $H(t,\FF_i)=0.2H(t,\FF_{i-1})+\eta_i$ for $t>0.5$.
	\item ($III_0^*$). $G(t,\FF_i)=H(t,\FF_i)\sqrt{1- (t-0.5)^2}/2$, where $H(t,\FF_i)=0.1H(t,\FF_{i-1})+\varepsilon_i$
	for $t\leq 0.5$, and $H(t,\FF_i)=0.4H(t,\FF_{i-1})+\eta_i$ for $t>0.5$,
\end{description}
where  $\FF_i=(\eta_{-\infty}\varepsilon_{-\infty}...,\eta_0, \varepsilon_0,...,\eta_i,\varepsilon_i)$.
In summary, the proposed test procedures work reasonably well as long as the underlying processes are not too different from PLS processes. An important class of non-stationary processes that are not PLS and cannot be handled by our methodology are the unit root non-stationary processes.
}

\begin{figure}[h]
\centering
   \includegraphics[width=14cm,height=10cm]{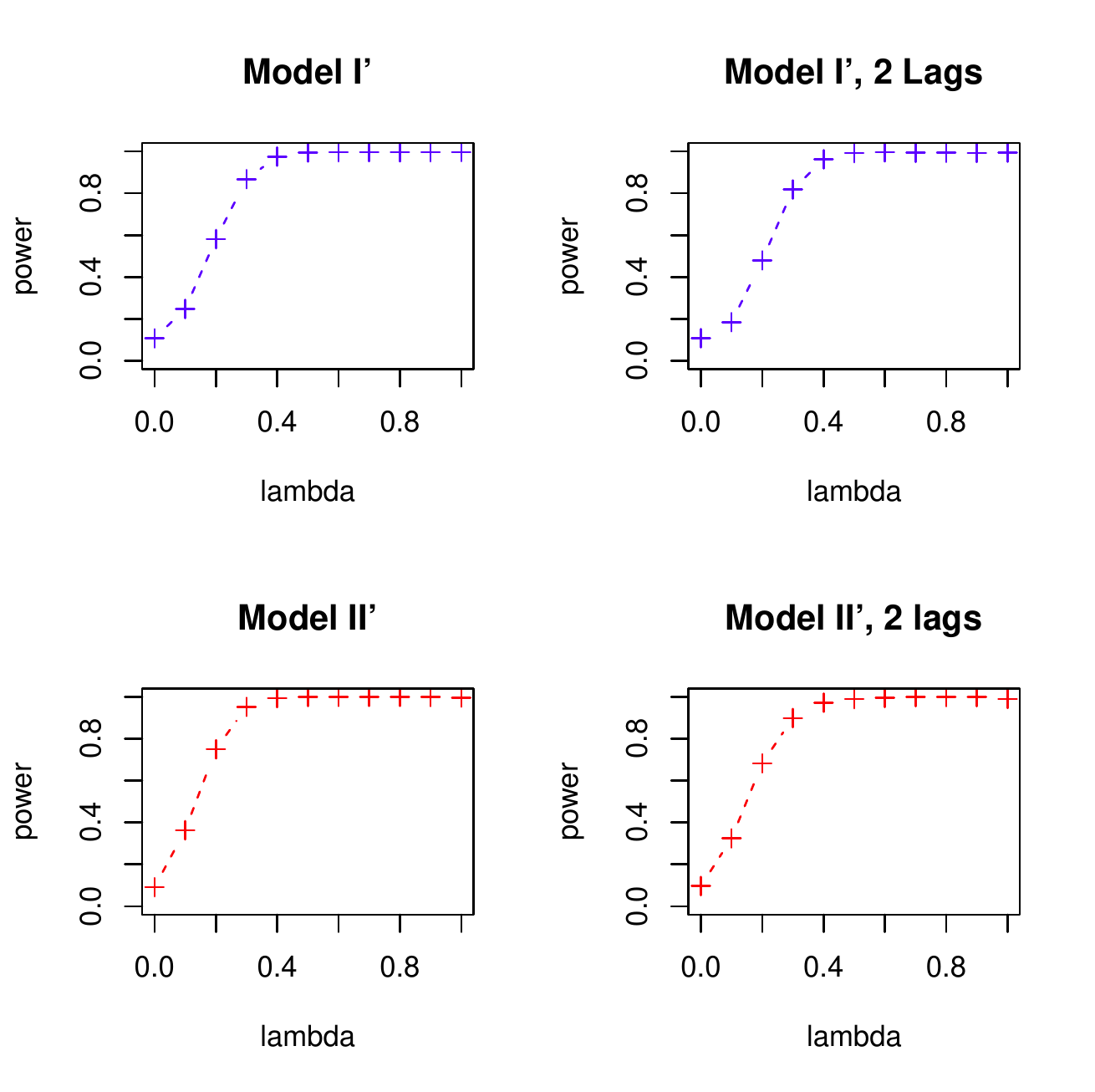}
\caption{\it Simulated power.
Upper left panel: test for a constant lag-$1$ correlation defined in  \eqref{testvar} (model  (I')).
Upper right panel: test for constant lags-$1$ and lag-$2$ correlation  defined in  \eqref{testvar} (model  (I')).
Lower left panel: test for the hypothesis of  a  non-relevant change in the lag-$1$ correlation defined in \eqref{bootrel}  (model (II')).
Lower right panel: test for the hypothesis of  a  non-relevant change in the lag-$1$ and lag-$2$ correlation  defined in \eqref{bootrel}  (model (II')).}
  \label{fig:mini:subfig:fige02}
\end{figure}

\subsection{Power properties}
In this section we investigate the power of the   proposed tests in two scenarios.
Let $\{\varepsilon_i\}_{i\in \mathbb Z}$ be $i.i.d$ N(0,1).
\begin{description}
\item (I') $G(t,\FF_i)=H(t,\FF_i)\sqrt{c(t)}/2$ for $t\leq 0.5$, and $G(t,\FF_i)=H_1(t,\FF_i)\sqrt{c(t)}/2$ for $t>0.5$, where
$c(t)=1- (t-0.5)^2$, $H(t,\FF_i)=0.2H(t,\FF_{i-1})+\varepsilon_i$ for $t\leq 0.5$, and $H(t,\FF_i)=(0.2-\lambda)H(t,\FF_{i-1})+\varepsilon_i$ for $t>0.5$.
\item (II') $G(t,\FF_i)=H(t,\FF_i)\sqrt{1- (t-0.5)^2}/2$, where $H(t,\FF_i)=(0.1-\lambda)H(t,\FF_{i-1})+\varepsilon_i$
for $t\leq 0.5$, and $H(t,\FF_i)=0.4H(t,\FF_{i-1})+\varepsilon_i$ for $t>0.5$.
\end{description}

{Model (I') is used to study the power of the test \eqref{testvar}
for the ``classical'' hypothesis of no change point in the correlation for various values of $\lambda$, where  $\lambda=0$ corresponds to the null hypothesis of a constant
correlation. In the upper panel  of Figure \ref{fig:mini:subfig:fige02} we show  the simulated power of the test for a constant  lag-$1$ correlation while in
 the upper right panel corresponding results of  the test for  a constant  lag-$1$  and lag-$2$ correlations are displayed.
{
We observe a decrease in power, which can be explained by the observation, that in  model (I')
the jump size of the lag $2$-correlation is $|0.2^2-(0.2-\lambda)^2|$, which is not monotone with respect to  $\lambda$.}
The power properties of the test \eqref{bootrel} of a  change  is investigated in model (II'). In the lower left panel of    Figure \ref{fig:mini:subfig:fige02}
 we  display the simulated rejection probabilities for
 the hypotheses of a non-relevant change in the  lag-$1$ correlation, that is
 $$H_0:\Delta_1\leq 0.3 ~\mbox{  versus } ~   H_1: \Delta_1 >0.3,$$
  where  $-0.6\leq\lambda\leq0$ corresponds to the null hypothesis. In
  the lower right panel  we investigate the hypotheses for lag-$1$  and lag-$2$ correlations, that is
  $$
  H_0:\Delta_1\leq 0.3\  \text{and}\  \Delta_2\leq 0.15~\mbox{  versus } ~ H_1:\Delta_1>0.3\  \text{or}\  \Delta_2 >0.15,
  $$
   where $0.1-\sqrt{0.31}\leq \lambda\leq0$ corresponds to the null hypothesis.
   {
We observe a decrease in power {(note again that, the jump size of the lag $2$-correlation is $|0.4^2-(0.1-\lambda)^2|$, which is not monotone with
respect to  $\lambda$).}
We conclude  that in all cases under consideration  the  proposed methodology can detect
(relevant) changes in the correlation structures with reasonable size.}

\begin{remark}{\rm \label{Remark-5.2-new}
{ 	The power of the proposed tests  depends sensitively on the choice of the bandwidth $b_n$. Ideally, if the errors are $i.i.d$ or
the series is strictly stationary, the optimal bandwidth can be calculated by an Edgeworth-expansion-based method (see \cite{Gao2008}) such that the power is optimized. However, the extension of this approach  to a PLS scenario is non-trivial, and  is out of the present  scope but an interesting problem for future work. In the case of a stationary null hypothesis, we have also compared the power of our test presented 
	with algorithms specifically designed for stationary processes. We observed  that our approach has decent power;
	these results  are presented in Section \ref{StationaryPerformance} of the supplementary material. }}
\end{remark}
\section{Data Analysis}\label{sec:data}
\def\theequation{6.\arabic{equation}}
\setcounter{equation}{0}
\begin{figure}[htbp]
\centering
  \includegraphics[width=15cm,height=10cm]{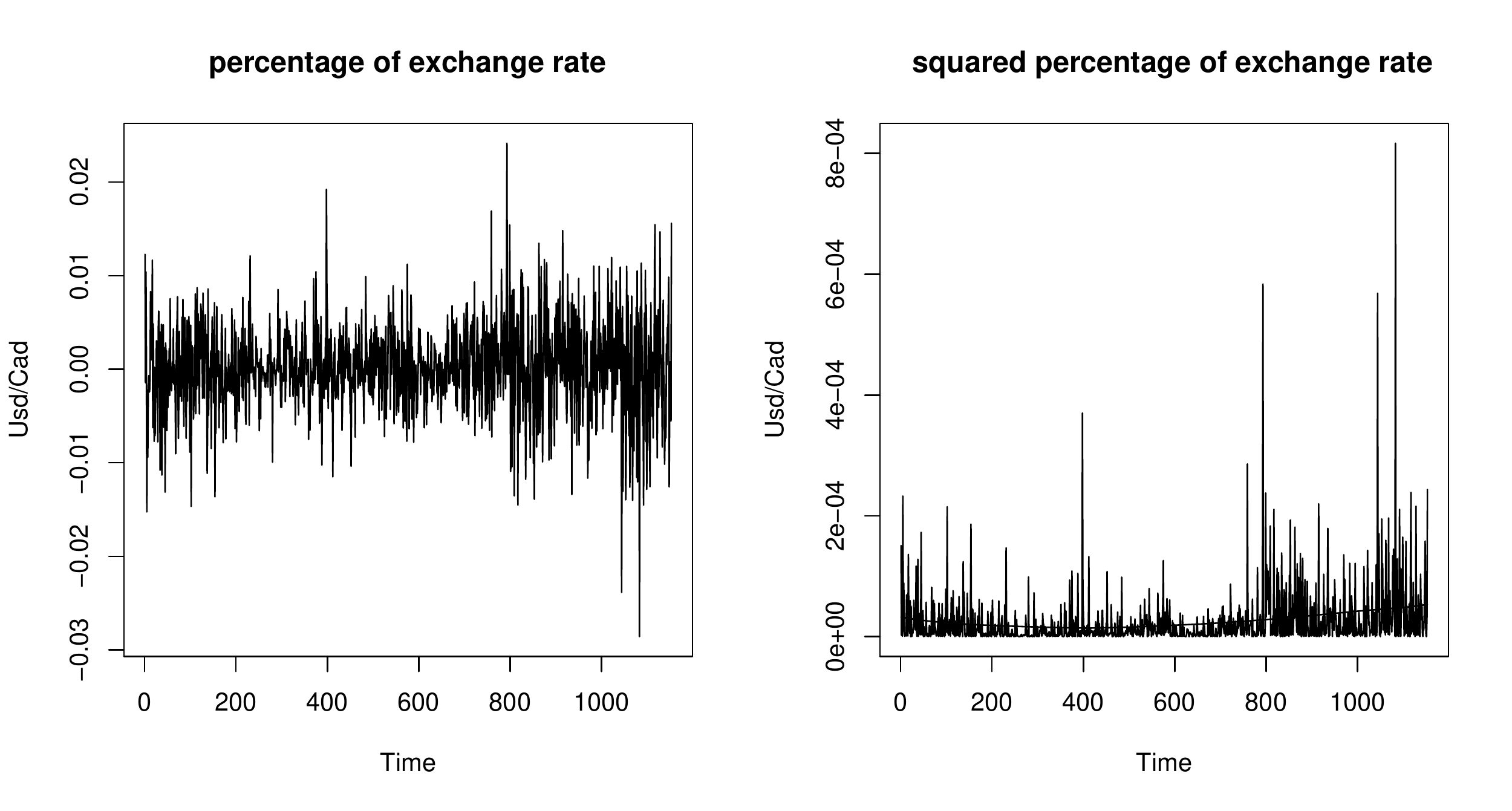}
 \caption{\it Percentage change (left panel) and the squared percentage change (right panel) of exchange rate of USD/CAD. The line in the right panel is the fitted mean for the squared percentage change.}
  \label{fig:mini:subfig:fige06}
\end{figure}
\begin{figure}[htbp]
\centering
  \includegraphics[width=15cm,height=10cm]{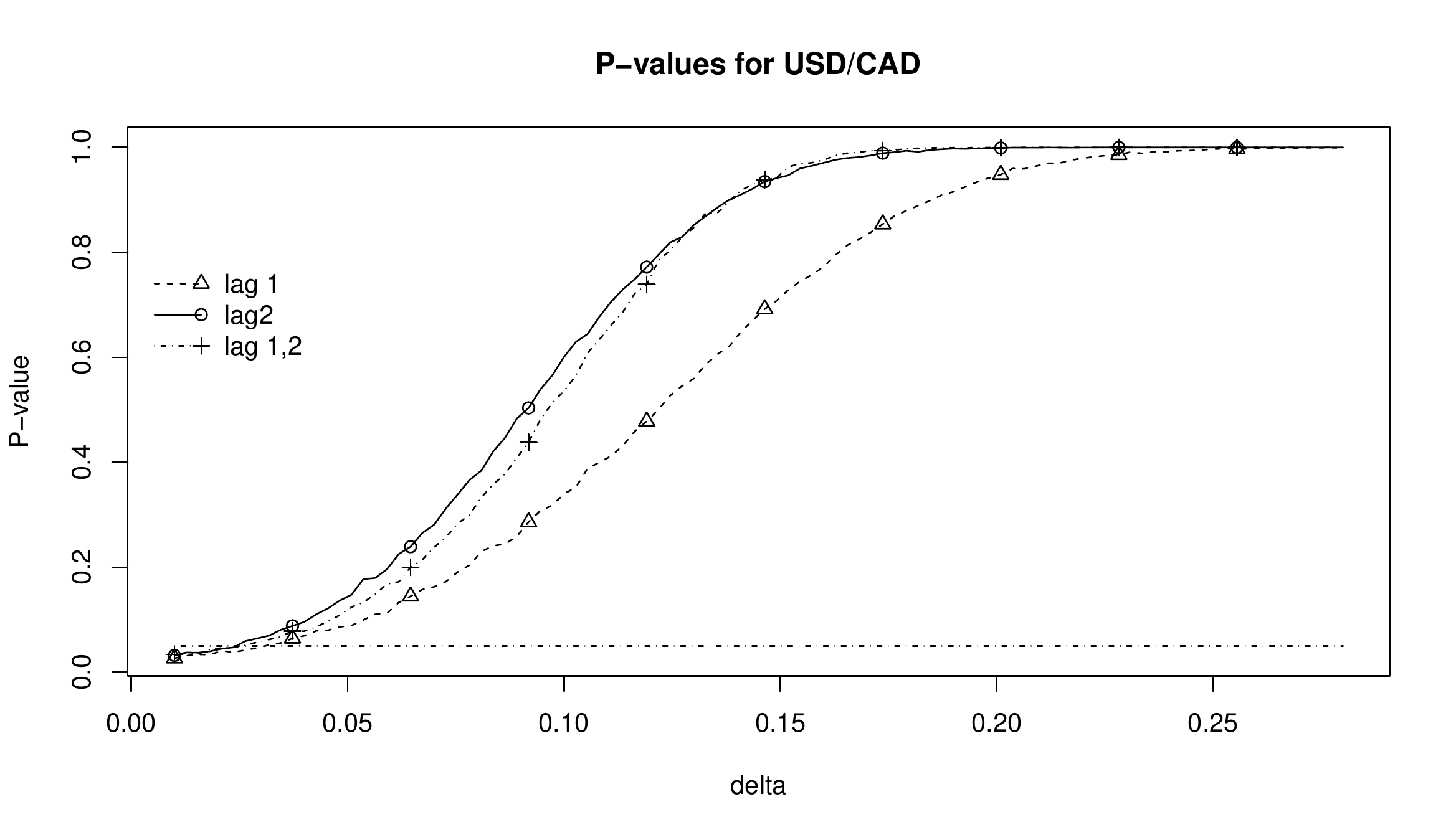}
 \caption{\it  p-values of the bootstrap test for a relevant change in  the lag-$1$, lag-$2$ and lag-($1$, $2$)  correlations for the squared percentage change of USD/CAD
 for different values of the threshold $\delta$. The horizontal line marks the significance
 level $0.05$.}
  \label{fig:mini:subfig:fige05}
\end{figure}

\begin{table}[htbp]
  \centering
  \caption{{\it Tests for the existence of a change point in the lag-$1$ and lag-$2$ correlations  of the USD/CAD series, respectively.
   $v^*_\alpha$ denotes the critical values obtained by the bootstrap procedure.
   ``Whole'' represents the whole period, ``Before'' and ``After'' represent the period before and after the detected change date.}}
    \begin{tabular}{|c|c|c|c|c|c|c|}
    \hline
          &  \multicolumn{3}{|c|}{lag $1$-Correlation}&  \multicolumn{3}{|c|}{lag $2$-Correlation} \\
    \hline
         & Whole\ \ \  & Before\ \ \  & After \ \ \   & Whole\ \ \  & Before\ \ \  & After \ \ \\
    \hline
    Test Stat.   & 1.28*  & 0.50  & 0.99    &1.38**  & 0.40    & 2.74\\
    $v^*_{90\%}$  & 1.23  & 0.70  & 1.01     &1.15    & 0.72    &3.95\\
    $v^*_{95\%}$  & 1.36  & 0.78  & 1.12     &1.3     &0.80     &4.70\\
    $b_n$       & 0.34    & 0.18    & 0.54       &0.34     &0.29    &0.15\\
    $m$        & 18        & 9    & 22         &18       &18      &12\\
    $c_n$       & 0.13    & 0.24    & 0.19      &0.13& 0.14 &0.34 \\
    \hline
    \end{tabular}%
 \label{tab:dataana-exchange}%
\end{table}%
We analyze the daily exchange rate of U.S. dollar/Canadian dollar from Nov 18th, 2011 to Jun 24th, 2016. The data can be obtained from https://www.federalreserve.gov/releases\\/h10/hist/. The series contains 1154 data points. During the period, USD/CAD has changed drastically in the range (0.9710, 1.4592). The wide range of the exchange rate motivates us to further investigate the robustness of the volatility of the percentage change of the series during the period. Figure \ref{fig:mini:subfig:fige06} shows the percentage change and squared percentage change of the exchange rate data. The pattern of the squared percentage change of exchange rate displays non-stationarity. For the test over the whole period, the GCV method selects $b_n=0.34$ and $c_n=0.3$ and
 the MV method select $m=18$. For this section, the critical values were generated by $8000$ bootstrap replications.  We used the statistic \eqref{tstar} to estimate the abrupt change points in the variance with $\zeta=0.10$ and $L=31$, and identified a variance change point  $t^*_n=795$, which corresponds to Jan 15th, 2015.

Let $X_t$ represent the squared percentage change at day $t$, and consider the relationship between $X_t$ and $X_{t-i}$, $i=1,2,3$. We performed our test on lag $1$, $2$ and $3$ simultaneously to check two null hypothesis: (i) all three correlations are $0$, (ii) all three correlations stay constant during the time considered. For (ii), we use the testing procedure in Section \ref{sec3}. For testing (i), we modified the test procedure in Section \ref{sec3} by
setting ${T}_n$ in \eqref{Tn} as ${T}_n=\max_{1\leq i\leq n}| {\mathbf S}^{}_i|$. The test statistic is the corresponding quantity $\hat T_n$ that replaces the error in $T_n$ by the local linear residuals. The critical value was generated by the bootstrap sample of $\max_{m+1\leq i\leq n-m+1}|{\hat {\mathbf \Phi}}_{i,m}|$ where $\hat {\mathbf \Phi}_{i,m}$ is defined in \eqref{new.83}.  For null hypothesis (i), the test statistics was 4.05, with simulated $p$-Value 1.2\%.
For null hypothesis (ii), the test statistics was 2.09 with simulated $p$-value 4.5\%. Hence there is moderately strong evidence that there are non-zero and non-constant correlations among the three lags.

 We analyzed the correlation at lags 1,2,3 separately. {\color{black}We tested the constancy in the lag-$1$ correlation for squared percentage change. 
  The $p$-value for the test of no change points in the lag-$1$ correlation was $7.6\%$. (see Table \ref{tab:dataana-exchange}), and the $p$-value for null hypothesis (i) of zero lag-1 correlation was $4.1\%$. }Next we used the statistic \eqref{changeestcorr} to identify the location of the change point of the first order correlation and found $\hat t_n=397$, which corresponds to Jun 18th, 2013. We investigated the existence of further changes in the lag-$1$ correlation before and after the Jun 18th, 2013 and concluded that there are no further structural breaks in the lag-$1$ correlation during the two periods at $5\%$ significance level, with the $p$-value $40\%$ and $11\% $, respectively, for the first and second period. For the first period, the test statistic for zero lag-$1$ correlation was $1.98$
   with $p$-value $<$ $1\%$. For the second period, the test statistic for zero lag-1 correlation was 2.25 with $p$-value $1.1\%$.
   {\color{black}
   The identified  change point  in the lag-$1$ correlation is close to the date that USD/CAD  significantly exceeded  the  boundary  $1$. Before this date, the exchange rate was slightly fluctuating  around $1$, and after
   this point the exchange rate increased over 1.4 and never returned to $1$.}

  For lag-2, the testing result are also presented in Table \ref{tab:dataana-exchange}. There $p$-value for the test of hypothesis of zero lag-2 correlation is $<1\% $, while the $p$-value of constant lag-2 correlation is $2.7\%$. The location of the jump time for the lag-2 correlation is 695 which corresponds to Aug 25, 2014. We also investigated the lag-2 correlation before and after Aug 25, 2014. For the hypothesis of constant lag-2 correlation,  The $p$-values were $71\%$ and $26\%$ for the first and second period, respectively. For the hypothesis of zero lag-2  correlation, the $p$-values were $1\%$ and $18\%$ before and after the jump, respectively. {\color{black}
  The  identified change point in the lag-$2$ correlation  is close to
   the date where  the Crude oil price drastically decreased from 100 USD per barrel to 50 USD per barrel. The oil price has a great impact on the economy of Canada, which is one of the decisive factors of the exchange rate.}

  For lag-3, the test statistic for no changes in correlation was 0.96 with $p$-value $43.3\%$. The test statistic for zero lag-3 correlation was  1.11 with $p$-value $58.9\%$. We conclude that correlations of the squared percentage changes concentrate in lag-1 and lag-2, with change points existing in both lags. Interestingly, the time of change for lag-1 and 2 correlations are different.

  We further performed tests from Section \ref{sec4} for relevant changes  
  in lag-1, lag-2 correlations separately and jointly (the trajectory we considered was $\delta_1=\delta_2$) for the USD/CAD data. The estimates of the lag-$1$ correlation before and after the break point were $-0.056$ and $0.079$ while, for the lag-2 correlation, the estimates before and after the jump were $0.092$ and $-0.034$.  The $p$-values of  the tests for a relevant change in the lag-$1$/lag-$2$ correlation for  different values of the threshold $\delta$ are displayed in Figure \ref{fig:mini:subfig:fige05}. At 5\% significance level, we conclude that there are relevant changes with size $\delta= 0.032$ in the lag-$1$ correlation, $\delta=0.024$ in the lag-2 correlation, and size $\delta=0.026$ in lag-1 or lag-2 correlation. 
 The $p$-values of the tests for relevant changes in the lag-$1$
  or $2$ correlation for different values of the threshold $\delta$ for the USD/CAD are displayed in Figure \ref{fig:mini:subfig:fige05}.

The correlations of the squared series are closely related to the ARCH effect. For example, \cite{BaillieChung2001} estimated the GARCH model via the autocorrelations of the square of the process. Our method shows that the USD/CAD from late 2011-mid 2016 may not be well fitted by a simple ARCH/GARCH model due to the changes in the correlation structure. Further, the negative first order correlation in the first period shows that USD/CAD from late 2011-mid 2013 may not be well fitted by usual ARCH/GARH model, due to their restriction of positive coefficients. Other models, for example the EGARCH model should be considered. We have also identified very different pattern of squared percentage changes of USD/CAD in the three lags considered: there is weak evidence against the null hypothesis of constant lag-1 correlation, strong evidence against constant lag-2 correlation, no evidence against constant lag-3 correlation, and strong evidence against constant lag-1, lag-2 and lag-3 correlations. We have no evidence against zero lag-3 correlation, while we have strong evidence against the hypotheses of zero lag-1 or lag-2 correlations.
\section{Supplementary Materials}
The supplementary materials contains the proofs of theorems and additional simulation results. 
\ \\\ \\
{\bf Acknowledgements.}
The work of H. Dette has been supported in part by the
Collaborative Research Center ``Statistical modeling of nonlinear
dynamic processes'' (SFB 823, Teilprojekt A1, C1) of the German Research Foundation (DFG). Z. Zhou's research has been supported in part by NSERC of Canada.
The authors would like to thank
Martina Stein who typed parts of this manuscript with considerable technical
expertise. The authors would also like to thank three referees and an associate editor for their constructive comments on an earlier version of this paper.

\bigskip
 \begin{small}

 \end{small}
\includepdf[pages=-]{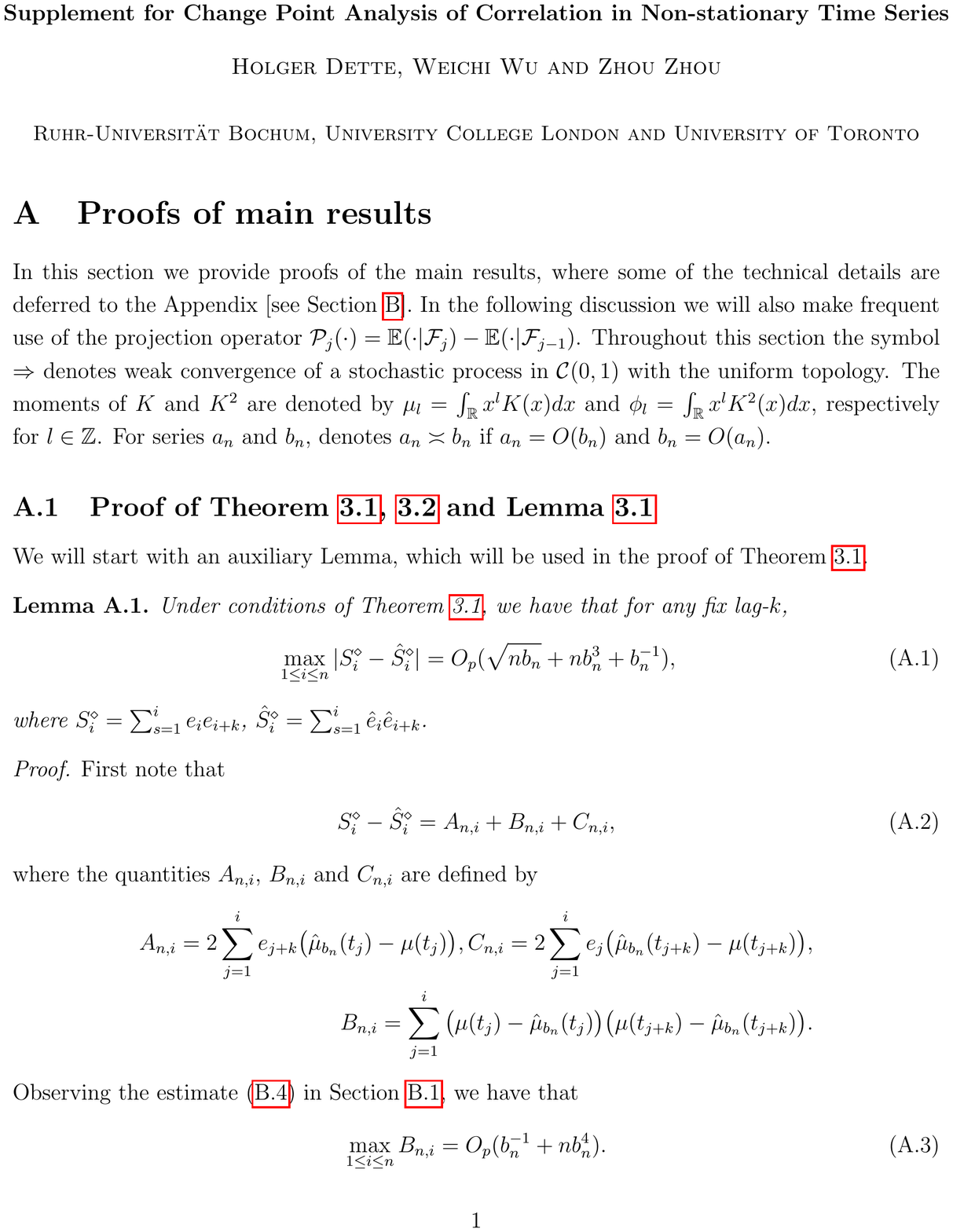}

\begin{thebibliography}{}



\bibitem[Abraham and Wei, 1984]{abrwei1984}
Abraham, B. and Wei, W. W.~S. (1984).
\newblock Inferences about the parameters of a time series model with changing
  variance.
\newblock {\em Metrika} 31,183--194.

\bibitem[Ahamada, Jouini and Boutahar, 2004]{ahajoubou2004}
Ahamada, I. and Jouini, J. and Boutahar, M.   (2004).
\newblock Detecting multiple breaks in time series covariance structure,
{A} nonparametric approach based on the evolutionary spectral density.
\newblock {\em Applied Economics}
  36,1095--1101.


\bibitem[Andrews, 1993]{andrews1993}
Andrews, D.~W.~K. (1993).
\newblock Tests for parameter instability and structural change with unknown change point.
\newblock {\em Econometrica} 61(4),128--156.

\bibitem[Aue et~al., 2009]{aueetal2009}
Aue, A., H{\"{o}}rmann, S., Horv{\'{a}}th, L., and Reimherr, M. (2009).
\newblock Break detection in the covariance structure of multivariate time
  series models.
\newblock {\em Annals of Statistics} 37(6),4046--4087.

\bibitem[Aue and Horv{\'{a}}th, 2013]{auehor2013}
Aue, A. and Horv{\'{a}}th, L. (2013).
\newblock Structural breaks in time series.
\newblock {\em Journal of Time Series Analysis} 34(1),1--16.

\bibitem[Baufays and Rasson, 1985]{bauras1985}
Baufays, P. and Rasson, J.~P. (1985).
\newblock Variance changes in autoregressive models.
\newblock In Anderson, D., editor, {\em Time Series Analysis, Theory and
  Practice 7} pages 119--127. North-Holland, New York.

\bibitem[Baillie and Chung, 2001]{BaillieChung2001}
Baillie, Richard T., and Huimin Chung. (2001).
\newblock Estimation of GARCH models from the autocorrelations of the squares of a process
\newblock {\em Journal of Time Series Analysis } 22(6),631--650.



\bibitem[Berger and Delampady, 1987]{bergdela1987}
Berger, J.~O. and Delampady, M. (1987).
\newblock Testing precise hypotheses.
\newblock {\em Statistical Science} 2(3),317--335.

\bibitem[Berkes  et~al., 2009]{bergomhor2009}
Berkes, I. and Gombay, E. and Horvath, L.   (2009).
\newblock Testing for changes in the co-variance structure of linear processes.
\newblock {\em Journal of Statistical Planning and Inference}
  139,2044--2063.

\bibitem[Berkson, 1938]{berkson1938}
Berkson, J. (1938).
\newblock Some difficulties of interpretation encountered in the application of
  the chi-square test.
\newblock {\em Journal of the American Statistical Association}
  33(203),526--536.

  \bibitem[Bollerslev, 1986]{Bollerslev1986}
  Bollerslev, T. (1986).
\newblock Generalized autoregressive conditional heteroskedasticity.
\newblock {\em Journal of econometrics}
  31(3), 307--327.

\bibitem[Chen and Gupta, 1997]{changevariance}
Chen, J. and Gupta, A.~K. (1997).
\newblock Testing and locating variance changepoints with application to stock
  prices.
\newblock {\em Journal of the American Statistical Association}
  92(438),739--747.

  \bibitem[Chow and Liu, 1992]{chowliu1992}
Chow, S.-C. and Liu, P.-J. (1992).
\newblock {\em Design and Analysis of Bioavailability and Bioequivalence Studies}.
\newblock {Marcel Dekker, New York}.

  \bibitem[Dahlhaus, 1997]{Dahlhaus1997}
Dahlhaus, R., (1997).
\newblock {Fitting time series models to nonstationary processes}.
\newblock {\em The Annals of Statistics} 25(1), 1-37.

\bibitem[Davis et~al., 2006]{davis2006}
Davis, R.~A., Lee, T. C.~M., and Rodriguez-Yam, G.~A. (2006).
\newblock Structural break estimation for nonstationary time series models.
\newblock {\em Journal of the American Statistical Association}
  101(473),223--239.


\bibitem[Dette  et~al., 2011]{detprevet2011}
  Dette, H. and Preu{{\ss}} P. and Vetter, M.  (2011).
\newblock A measure of stationarity in locally stationary processes with applications to testing.
\newblock {\em Journal of the American Statistical Association}
  106,1113--1124.

  \bibitem[Dette and Wied, 2016]{dettwied2016}
Dette, H. and Wied, D. (2016).
\newblock Detecting relevant changes in time series models.
\newblock {\em Journal of the Royal Statistical Society, Ser., B}
  78,371--394.

\bibitem[Dette et~al., 2015]{holger2015}
Dette, H., Wu, W. and Zhou, Z. (2015).
\newblock Change point analysis of second order characteristics in non-stationary time series.
\newblock {	arXiv,1503.08610 }.


\bibitem[Dwivedi and Subba Rao, 2011]{dwisub2011}
Dwivedi, Y. and Subba Rao, S. (2011).
\newblock A test for second order stationarity of a time series based on the discrete  {F}ourier transform.
\newblock {\em Journal of Time Series Analysis}
  32,68--91.

  \bibitem[Engle,  1982]{Engle1982}
Engle, Robert F. (1982)
\newblock Autoregressive conditional heteroscedasticity with estimates of the variance of United Kingdom inflation.
\newblock {\em Econometrica} (1982), 987--1007.

  \bibitem[Fan and Gijbels, 1996]{fangij1996}
Fan, J. and Gijbels, I. (1996).
\newblock {\em Local Polynomial Modelling and its Applications}.
\newblock {Chapman \& Hall, London}.

\bibitem[Galeano and Pe{\~{n}}a, 2007]{galpen2007}
Galeano, P. and Pe{\~{n}}a, D. (2007).
\newblock Covariance changes detection in multivariate time series.
\newblock {\em Journal of Statistical Planning and Inference} 137,194--211.

\bibitem[Gao and Gijbels, 2008]{Gao2008}
Gao, J. and  Gijbels, I. (2008).
\newblock Bandwidth Selection in Nonparametric Kernel Testing
\newblock  {\em Journal of the American Statistical Association} {\bf 103} 1584--1594.

\bibitem[Tong, 1990]{Tong1990}
Tong, H. (1990).
\newblock Non-linear time series, a dynamical system approach
\newblock {\em  Oxford , Oxford University Press}.


\bibitem[Incl{\'{a}}n and Tiao, 1994]{incltiao1994}
Incl{\'{a}}n, C. and Tiao, G.~C. (1994).
\newblock Use of cumulative sums of squares for retrospective detection of
  changes of variance.
\newblock {\em Journal of the American Statistical Association} 89(427).

\bibitem[Jandhyala et~al., 2013]{jandhyala2013}
Jandhyala, V., Fotopoulos, S., MacNeill, I., and Liu, P. (2013).
\newblock Inference for single and multiple change-points in time series.
\newblock {\em Journal of Time Series Analysis} 34(4),423--446.


\bibitem[Jin et~al.,  2015]{jinwang2015}
  Jin, L. and Wang, S. and Wang, H. (2015).
\newblock A new non-parametric stationarity test of time series in the time domain.
\newblock {\em Journal of the Royal Statistical Society, Series B}
  77,893--922.



\bibitem[Killick  et~al., 2013]{kileckjon2013}
Killick, R. and Eckley, I.~A. and Jonathan, P.   (2013).
\newblock A wavelet-based approach for detecting changes in second order structure within nonstationary time series.
\newblock {\em Electronic Journal of Statistics}
  7,1167--1183.

\bibitem[Lee and Park, 2001]{leepark2001}
Lee, S. and Park, S. (2001).
\newblock The cusum of squares test for scale changes in infinite order moving
  average processes.
\newblock {\em Scandinavian Journal of Statistics} 28(4),625--44.

\bibitem[Mallik et~al., 2013]{Mallik2013}
Mallik, A., Banerjee, M., and Sen, B. (2013).
\newblock Asymptotics for {$p$}-value based threshold estimation in regression
  settings.
\newblock {\em Electronic Journal of Statistics} 7,2477--2515.


\bibitem[Mallik et~al., 2011]{Mallik2011}
Mallik, A., Sen, B., Banerjee, M., and Michailidis, G. (2011).
\newblock Threshold estimation based on a p-value framework in dose-response
  and regression settings.
\newblock {\em Biometrika} 98,887--900.

\bibitem[McBride, 1999]{mcbride1999}
Mc{B}ride, G.~B. (1999).
\newblock Equivalence tests can enhance environmental science and management.
\newblock {\em Australian {$\&$} New Zealand Journal of Statistics} 41,19--29.

\bibitem[M{\"u}ller, 1992]{Mueller1992}
M{\"u}ller, H.-G. (1992).
\newblock Change-points in nonparametric regression analysis.
\newblock {\em Annals of Statistics} 20,737--761.

\bibitem[Page, 1954]{page1954}
Page, E.~S. (1954).
\newblock Continuous inspection schemes.
\newblock {\em Biometrika} 41,100--115.


\bibitem[Paparoditis, 2009]{paparoditis2009}
Paparoditis, E. (2009).
\newblock Testing temporal constancy of the spectral structure of a time series.
\newblock {\em Bernoulli}
  15,1190--1221.

  \bibitem[Paparoditis, 2010]{paparoditis2010}
Paparoditis, E. (2010).
\newblock Validating stationarity assumptions in time series analysis by rolling local periodograms.
\newblock {\em Journal of the American Statistical Association}
  105,839--851.

\bibitem[Paparoditis and Preu{{\ss}}, 2015]{papapreu2015}
Paparoditis, E. and Preu{{\ss}} P.  (2015).
\newblock  On local power properties of frequency domain-based tests for stationarity.
\newblock {\em Scandinavian Journal of Statistics}
43,664-682.

\bibitem[Politis et~al., 1999]{polromwol1999}
Politis, D.~N., Romano, J.~P., and Wolf, M. (1999).
\newblock {\em Subsampling}.
\newblock Springer, New York.


  \bibitem[Preu{{\ss}} et~al., 2013]{prevetdet2013}
Preu{{\ss}} P. and Vetter, M. and Dette, H. (2013).
\newblock A test for stationarity based on empirical processes.
\newblock {\em Bernoulli}
  19,2715--2749.


\bibitem[Preuss et~al., 2015]{prepucdet2015}
Preuss, P., Puchstein, R., and Dette, H. (2015).
\newblock Detection of multiple structural breaks in multivariate time series.
\newblock {\em Journal of the American Statistical Association}
110,654-668.


\bibitem[Qu, 2008]{qu2008}
Qu, Z. (2008).
\newblock Testing for structural change in regression quantiles.
\newblock {\em Journal of Econometrics} 148,170--184.


\bibitem[Solomon, 2007]{solomon2007}
Solomon, S. (2007).
\newblock {\em Climate change 2007-the physical science basis, Working group I
  contribution to the fourth assessment report of the IPCC} volume~4.
\newblock Cambridge University Press.

\bibitem[Vogt, 2012]{vogt2012}
Vogt, M. (2012).
\newblock Nonparametric regression for locally stationary time series.
\newblock {\em The Annals of Statistics} 40(5),2601--2633.

\bibitem[Vogt and Dette, 2015]{vogtdett2015}
Vogt, M. and Dette, H. (2015).
\newblock Detecting gradual changes in locally stationary processes.
\newblock {\em Annals of Statistics} 43(2),713--740.

\bibitem[Vostrikova, 1981]{vostrikova1981}
Vostrikova, L.~J. (1981).
\newblock Detecting `disorder' in multidimensional random processes.
\newblock {\em Soviet Mathematics Doklady}
  24,55--59.

\bibitem[Wichern et~al., 1976]{wichern1976}
Wichern, D.~W., Miller, R.~B., and Hsu, D.-A. (1976).
\newblock Changes of variance in first-order autoregressive time series models
  - with an application.
\newblock {\em Journal of the Royal Statistical Society. Ser. C (Applied
  Statistics)} 25(3),248--256.

\bibitem[Wied et~al., 2012]{wiekradeh2012}
Wied, D., Kr{\"{a}}mer, W., and Dehling, H. (2012).
\newblock Testing for a change in correlation at an unknown point in time using
  an extended functional delta method.
\newblock {\em Econometric Theory} 28(3),570--589.

\bibitem[Wu, 2005]{wu2005}
Wu, W.~B. (2005).
\newblock Nonlinear system theory, {A}nother look at dependence.
\newblock {\em Proceedings of the National Academy of Sciences of the United
  States of America} 102(40),14150--14154.
\bibitem[Zhang and Wu, 2012]{zhang2012}
Zhang, T., and Wu, W. B. (2012).
\newblock Inference of time-varying regression models.
\newblock {\em The Annals of Statistics} 40(3), 1376--1402.

\bibitem[Zhou and Wu, 2009]{zhouwu2009}
Zhou, Z. and Wu, W.B. (2009).
\newblock Local linear quantile estimation for nonstationary time series.
\newblock {\em Annals of Statistics} 2696-2729.


\bibitem[Zhou and Wu, 2010]{zhouwu2010}
Zhou, Z. and Wu, W.~B. (2010).
\newblock Simultaneous inference of linear models with time-varying
  coefficients.
\newblock {\em Journal of the Royal Statistical Society, Series B}
  72,513--531.

\bibitem[Zhou, 2012]{Zhou2012}
Zhou, Z (2012).
\newblock Measuring Nonlinear Dependence in Multivariate Time Series, A Distance Correlation Approach.
\newblock {\em Journal of Time Series Analysis} 33, 438--457.


\bibitem[Zhou, 2013]{zhou2013}
Zhou, Z. (2013).
\newblock Heteroscedasticity and autocorrelation robust structural change
  detection.
\newblock {\em Journal of the American Statistical Association} 108,726--740.

\bibitem[Zhou, 2014]{zhou2014}
Zhou, Z. (2014).
\newblock Inference of weighted {$V$}-statistics for non-stationary time series
  and its applications.
\newblock {\em Annals of Statistics} 42,87--114.


\end{thebibliography}
\end{document}